\documentclass[twocolumn,english,amssymb,prl,superscriptaddress,notitlepage,nolongbibliography]{revtex4-2}
\usepackage{lmodern}
\usepackage{lmodern}
\usepackage[T1]{fontenc}
\usepackage[latin9]{inputenc}
\usepackage{color}
\usepackage{babel}
\usepackage{bm}
\usepackage{amsmath}
\usepackage{amssymb}
\usepackage{graphicx}
\usepackage{esint}
\usepackage[unicode=true,pdfusetitle,
 bookmarks=true,bookmarksnumbered=true,bookmarksopen=false,
 breaklinks=true,pdfborder={0 0 1},backref=false,colorlinks=true]
 {hyperref}
\hypersetup{breaklinks,
 linkcolor=magenta,urlcolor=blue,citecolor=blue,pdfstartview={FitH},hyperfootnotes=false}

\makeatletter
\usepackage{babel}
\date{\today}
\usepackage{xcolor}
\allowdisplaybreaks

\def\Bigr{\mathclose\Big}

\renewcommand{\big}{\bBigg@\@ne}
\renewcommand{\Big}{\bBigg@{1.5}}
\renewcommand{\bigg}{\bBigg@\tw@}
\renewcommand{\Bigg}{\bBigg@{2.5}}

\newcommand{\biggg}{\bBigg@\thr@@}
\newcommand{\Biggg}{\bBigg@{3.5}}

\makeatother

\begin{document}
\title{Topological Multiband $s$-wave Superconductivity in Coupled Multifold
Fermions}
\author{Changhee Lee}
\affiliation{Department of Physics and Astronomy, Seoul National University, Seoul
08826, Korea}
\author{Chiho Yoon}
\affiliation{Department of Physics and Astronomy, Seoul National University, Seoul
08826, Korea}
\author{Taehyeok Kim}
\affiliation{Department of Physics and Astronomy, Seoul National University, Seoul
08826, Korea}
\author{Suk Bum Chung}
\email{sbchung0@uos.ac.kr}

\affiliation{Department of Physics, University of Seoul, Seoul 02504, Korea}
\affiliation{Natural Science Research Institute, University of Seoul, Seoul 02504,
Korea}
\affiliation{School of Physics, Korea Institute for Advanced Study, Seoul 02455,
Korea}
\author{Hongki Min}
\email{hmin@snu.ac.kr}

\affiliation{Department of Physics and Astronomy, Seoul National University, Seoul
08826, Korea}
\begin{abstract}
We study three-dimensional time-reversal-invariant topological superconductivity
in noncentrosymmetric materials such as RhSi, CoSi, and AlPt which
host coupled multifold nodes energetically split by the spin-orbit
coupling at the same time-reversal-invariant momentum (TRIM). The
topological superconductivity arises from the $s_{+}\oplus s_{-}$
gap function, which is ${\bm{k}}$ independent, but with opposite
signs for the two nodes split at the same TRIM. We consider various
electron-electron interactions in the tight-binding model for RhSi
and find that the topological superconducting phase supporting a surface
Majorana cone and topological nodal rings is favored in a wide range
of interaction parameters. 
\end{abstract}
\maketitle
\textit{Introduction.}--- Majorana particles on superconductors have
drawn widespread attention as an avenue to topological quantum computation
\citep{Alicea2012,Beenakker2013,Stanescu2013,Sato2016,Sato2017},
and this can be attributed to their characteristic non-abelian vortex
exchange statistics \citep{Read2000,Fidkowski2013,Sanno2021}. The majority of experimental search for the Majorana zero modes (MZMs)
in condensed matter physics has been based on proposals to realize
them in a heterostructure. Examples include the semiconductor nanowire~\citep{Oreg2010,Sau2010,Lutchyn2010,Liu2017,Prada2020},
the ferromagnetic atomic chain~\citep{NadjPerge2014},
and the topological insulator~\citep{Fu2008,Fu2009} in proximity to a conventional superconductor. 

Besides these heterostructures, the development of topological
band theory \citep{Fu2007,Qi2010,Chiu2016} has demonstrated that
topological materials could exhibit topological superconductivity (TSC) with Majorana boundary modes due to the novel interplay between
the unique electronic structures and interactions \citep{Moon2013,Herbut2014,Maciejko2014,Detassis2017,Han2019,Shi2021}.
Recent research has been devoted to understanding the interplay and
the resulting TSC in Weyl/Dirac semimetals
\citep{XWan2011, Burkov2011, Wehling2014,Armitage2018} and Luttinger
semimetals \citep{Murakami2004,TKondo2015}.

The recently discovered topological semimetals with unconventional
multifold fermions \citep{Bradlyn2016} also provide an ideal platform
for studying the effects of interactions in the unique electronic
structures, especially TSC. In a family
of candidate materials with the B20 crystal structure \citep{Muhlbauer2009,Pshenay-Severin2019}
such as RhSi, CoSi and AlPt, the topological features of the electronic
structure such as the long surface Fermi arcs have been observed \citep{Chang2017,Ni2020,Rees2020,Sanchez2019,Takane2019,Rao2019,Schroter2019}.
However, some qualitative band structure features of this class of
candidate materials have not been included in the existing studies
on the interacting multifold fermions \citep{Isobe2016,Link2020,Boettcher2020,Huang2020,Gao2020}.
Many of them studied isolated multifold fermions at a single time-reversal
invariant momentum (TRIM) and it has not been considered that two
types of multifold fermions lie at slightly different energy levels
on the same TRIM. Thus, the possibility of the multi-band characteristics
to manifest in the superconductivity \citep{Dai2008,Namoto2016,Suh2019,Setty2020}
has been overlooked.

In this Letter, we study a possible time-reversal invariant (TRI)
TSC arising from the coupled structure of multifold fermions
at TRIMs in a representative multifold fermion system, RhSi. We find
that 
the electronic structure of coupled multifold fermions can give rise to a robust combination of
the fully gapped TRI TSC around the coupled nodes at $\Gamma$ and
the nodal TSC around another TRIM, which
arises from the $\bm{k}$-independent intra-nodal pairing gap at each
TRIM with opposite signs between the two nodes at the same TRIM; we
shall call this the ``$s_{+}\oplus s_{-}$ gap function''.

\begin{figure}
\includegraphics{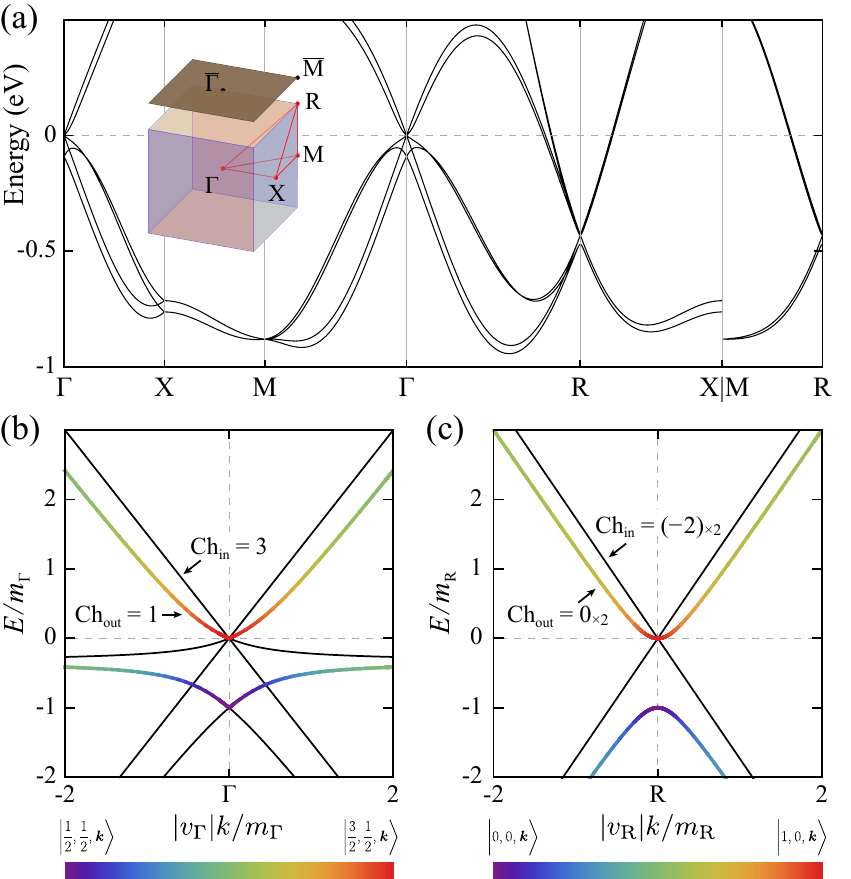} \caption{\label{fig1}(a) Band structure of RhSi along the high-symmetry lines
in the BZ. The BZ and the (001) SBZ are shown in the inset. (b),(c)
The band structure of the continuum model near the nodal points at
$\Gamma$ (${\rm R}$). The ratio of the wave function magnitude between
$|\frac{3}{2},\frac{1}{2},\hat{\bm{k}}\rangle$ and $|\frac{1}{2},\frac{1}{2},\hat{\bm{k}}\rangle$
($|1,0,\hat{\bm{k}}\rangle$ and $|0,0,\hat{\bm{k}}\rangle$) is illustrated
by different colors. For the bands with $E>0$, ${\rm Ch}_{{\rm in}}$
and ${\rm Ch}_{{\rm out}}$ are the Chern numbers carried by the inner
and outer FSSs, respectively, and $\times2$ in the subscript indicates
double degeneracy.}
\end{figure}

\textit{Low energy effective model.}--- To study the electronic structure
of RhSi, we adopt the tight-binding (TB) model proposed in Ref.~\citep{Chang2017}
(see Sec. I of Supplemental Material (SM) \citep{SM} for the details).
Figure~\ref{fig1}(a) shows the TB band structure of RhSi and the
first Brillouin zone (BZ) along with the surface BZ (SBZ) on the (001)
plane. At $\Gamma$, four-fold and two-fold nodal points are found
around the zero energy corresponding to the (pseudo)spin-$\frac{3}{2}$
Rarita-Schwinger-Weyl (RSW) fermion and spin-$\frac{1}{2}$ Kramer-Weyl
(KW) fermion, respectively. Additionally, there are six-fold and two-fold
nodal points at ${\rm R}$ which are equivalent to doubly degenerate
spin-$1$ \citep{Takane2019,Hu2018,Fulga2017,Lin2018} and spin-$0$
systems, respectively. As shown in Fig.~\ref{fig1}(a), the Fermi
level of RhSi lies around the four-fold nodal point at $\Gamma$ and
the energy bands crossing the Fermi level emanate from the multifold
nodal points at $\Gamma$ and ${\rm R}$. 

We first investigate a low-energy effective model around $\Gamma$.
Up to first order in $\boldsymbol{k}$, the electronic band structure
near the two nodal points at $\Gamma$ is described by the following
Hamiltonian:
\begin{equation}
H_{\bm{k}}^{\frac{3}{2}\oplus\frac{1}{2}}=\left(\begin{matrix}H_{\bm{k}}^{(\frac{3}{2})} & T_{\Gamma,\bm{k}}\\
T_{\Gamma,\bm{k}}^{\dagger} & H_{\bm{k}}^{(\frac{1}{2})}-m_{\Gamma}I_{2}
\end{matrix}\right),\label{eq:H_Gamma}
\end{equation}
where $H_{\bm{k}}^{(\frac{3}{2})}=v_{\Gamma}\bm{k}\cdot\bm{S}^{(\frac{3}{2})}$
and $H_{\bm{k}}^{(\frac{1}{2})}=2v_{\Gamma}\bm{k}\cdot\bm{S}^{(\frac{1}{2})}$
describe the RSW and KW fermions, respectively, with the spin-$s$
matrices $S_{i}^{(s)}$ satisfying $\bm{S}^{(s)2}=s(s+1)$. Here,
$v_{\Gamma}$ and $2v_{\Gamma}$ are the Fermi velocities of RSW and
KW fermions, respectively. In RhSi family, they originate from spin-independent
inversion-breaking hoppings. We denote the eigenstates of
$H_{\boldsymbol{k}}^{(s)}$ by $|s,h_{z},\hat{\bm{k}}\rangle$, where
$s$ and $h_{z}$ are the total angular momentum and the helicity
eigenvalue, respectively. We assume $v_{\Gamma}>0$ without loss of
generality. The two nodal points are energetically separated by $m_{\Gamma}$
and coupled by $T_{\Gamma,\boldsymbol{k}}=\frac{v_{\Gamma}k}{\sqrt{2}}\big(|\frac{3}{2},\frac{1}{2},\hat{\bm{k}}\rangle\langle\frac{1}{2},\frac{1}{2},\hat{\boldsymbol{k}}|+|\frac{3}{2},-\frac{1}{2},\hat{\bm{k}}\rangle\langle\frac{1}{2},-\frac{1}{2},\hat{\boldsymbol{k}}|\big)$.

When the Fermi level lies above the four-fold nodal point, two bands
of $H_{\boldsymbol{k}}^{\frac{3}{2}\oplus\frac{1}{2}}$ cross the Fermi level as shown in Fig.~\ref{fig1}(b). Their eigenstates are
given by 
\begin{equation}
\begin{split}|\Gamma,\text{in},\bm{k}\rangle & =\left|\frac{3}{2},\frac{3}{2},\hat{\bm{k}}\right\rangle ,\\
|\Gamma,\text{out},\bm{k}\rangle & =\cos\theta_{\Gamma,k}\left|\frac{3}{2},\frac{1}{2},\hat{\bm{k}}\right\rangle +\sin\theta_{\Gamma,k}\left|\frac{1}{2},\frac{1}{2},\hat{\bm{k}}\right\rangle ,
\end{split}
\label{eq:H_Gamma_eigenstate}
\end{equation}
where $\theta_{\Gamma,k}=\tan^{-1}\sqrt{\frac{f_{\Gamma,k}-e_{\Gamma,k}}{f_{\Gamma,k}+e_{\Gamma,k}}}$
with $e_{\Gamma,k}=m_{\Gamma}-v_{\Gamma}k/2$ and $f_{\Gamma,k}=\sqrt{e_{\Gamma,k}^{2}+2v_{\Gamma}^{2}k^{2}}$.
Here, ``in'' and ``out'' represent the inner and outer Fermi surface
sheets (FSSs), respectively. Note that the outer FSS mainly consists
of the $h_{z}=\frac{1}{2}$ branch of the RSW fermion in the limit
$v_{\Gamma}k\ll m_{\Gamma}$, while the portion of the KW fermion
$|\frac{1}{2},\frac{1}{2},\hat{\boldsymbol{k}}\rangle$ increases
to $\frac{2}{3}$ in the opposite limit. In Fig.~\ref{fig1}(b),
the bands composed of the $h_{z}=\frac{1}{2}$ branches of the RSW
and KW fermions are colored according to their ratio. A FSS surrounding
a multifold fermion carries a quantized Chern number $2h_{z}$ when
each eigenstate on the FSS is a superposition of $|s,h_{z},\hat{\boldsymbol{k}}\rangle$'s
with the same $h_{z}$. Thus, the inner and outer FSSs carry the Chern
numbers 3 and 1, respectively.

The low-energy effective model near ${\rm R}$ is analogous to that
near $\Gamma$. Up to first order in $\bm{k}$, the electronic band
structure around six-fold and two-fold nodal points at ${\rm R}$
is described by two copies of coupled spin-1 and spin-0 fermions $H_{\boldsymbol{k}}^{1\oplus0}\oplus H_{\boldsymbol{k}}^{1\oplus0}$,
where $H_{\boldsymbol{k}}^{1\oplus0}$ is given by 
\begin{equation}
H_{\bm{k}}^{1\oplus0}=\left(\begin{matrix}H_{\bm{k}}^{(1)} & T_{{\rm R},\bm{k}}\\
T_{{\rm R},\bm{k}}^{\dagger} & H_{\bm{k}}^{(0)}
\end{matrix}\right)=\left(\begin{matrix}-v_{{\rm R}}\bm{k}\cdot\bm{S}^{(1)} & T_{{\rm R},\bm{k}}\\
T_{{\rm R},\bm{k}}^{\dagger} & -m_{{\rm R}}
\end{matrix}\right),\label{eq:H_R}
\end{equation}
where the two nodal points are coupled by $T_{\mathrm{R},\bm{k}}=v_{\mathrm{R}}k|1,0,\hat{\boldsymbol{k}}\rangle\langle0,0,\hat{\boldsymbol{k}}|$.
Here, $v_{\text{R}}>0$ since the Nielsen-Ninomiya theorem~\citep{Nielsen1981}
requires $v_{\Gamma}v_{\text{R}}>0$.

As shown in Fig.~\ref{fig1}(c), two bands of $H_{\bm{k}}^{1\oplus0}$
cross the Fermi level when it lies above the six-fold nodal point,
whose eigenstates are given by 
\begin{equation}
\begin{split}|\mathrm{R},\text{in},\bm{k}\rangle & =|1,-1,\hat{\bm{k}}\rangle,\\
|\mathrm{R},\text{out},\bm{k}\rangle & =\cos\theta_{\text{R},k}|1,0,\hat{\bm{k}}\rangle+\sin\theta_{\text{R},k}|0,0,\hat{\bm{k}}\rangle,
\end{split}
\label{eq:H_R_eigenstate}
\end{equation}
where $\theta_{\text{R,}k}=\tan^{-1}\left(\sqrt{\frac{f_{\mathrm{R},k}-m_{{\rm R}}}{f_{\mathrm{R},k}+m_{{\rm R}}}}\right)$
with $f_{\mathrm{R},k}=\sqrt{m_{{\rm R}}^{2}+4v_{{\rm R}}^{2}k^{2}}$.
Eq.~\eqref{eq:H_R_eigenstate} is very similar to Eq.~\eqref{eq:H_Gamma_eigenstate}.
The Chern numbers carried by the
inner and outer FSSs are $-2$ and 0, respectively, for each copy.
Then the total Chern number of the Fermi surface around ${\rm R}$
is $-4$ compensating the Chern numbers from the FSSs around the RSW
and KW nodes at $\Gamma$.

\textit{TSC from $s_{+}\oplus s_{-}$ pairings.}--- Considering the
two coupled nodal points at $\Gamma$ or ${\rm R}$, TRI
TSC can arise even from the simplest $\bm{k}$-independent
gap functions represented by a direct sum of two trivial matrices
respecting all the spatial symmetries of the system. Besides the trivial gap function, the multinodal nature of each TRIM in the system
allows the $s_{+}\oplus s_{-}$ gap functions: 
\begin{equation}
\Delta_{\pm}^{\Gamma}=\left(\begin{matrix}I_{4}\\
 & -2I_{2}
\end{matrix}\right),\quad\Delta_{\pm}^{{\rm R}}=\left(\begin{matrix}I_{3}\\
 & -3I_{1}
\end{matrix}\right),\label{eq:Delta_pm}
\end{equation}
where $I_{2s+1}$ acts on spin-$s$ fermions. $\Delta_{\pm}^{\Gamma(\text{R})}$
looks like the trivial gap function if we just focus on one
of the two nodal points at $\Gamma$ (${\rm R}$), but the sign of
the gap function on each nodal point is opposite. Note that the $s_{+}\oplus s_{-}$
gap functions indicate spin-triplet pairing (see Sec. II of SM \citep{SM}).
We shall show later that, for generic electron-electron interactions,
the TRI solution of the linearized gap equation can be written as
\begin{equation}
\begin{split}\Delta_{\Gamma}(\beta_{\Gamma}) & =\alpha_{\Gamma}\left(\Delta_{\pm}^{\Gamma}\cos\beta_{\Gamma}+I_{6}\sin\beta_{\Gamma}\right),\\
\Delta_{{\rm R}}(\beta_{\mathrm{R}}) & =\alpha_{{\rm R}}\left(\Delta_{\pm}^{{\rm R}}\cos\beta_{{\rm R}}+I_{4}\sin\beta_{{\rm R}}\right),
\end{split}
\label{eq:Delta_parameterization}
\end{equation}
near $\Gamma$ and ${\rm R}$, respectively. Here, $\alpha_{\Gamma(\text{R})}$
corresponds to the overall magnitude of the gap function and $\beta_{\Gamma(\text{R})}\in[-\frac{\pi}{2},\frac{\pi}{2}]$
parameterizes the ratio between the trivial gap function and the $s_{+}\oplus s_{-}$
gap function. We assume $\alpha_{\Gamma},\alpha_{\text{R}}>0$
here (see Sec. III of SM \citep{SM} for more general cases).

To understand the topological nature of the superconducting phase
with the gap function $\Delta_{\Gamma}(\beta_{\Gamma})$, we first
focus on the multifold fermions at $\Gamma$ and study the Bogliubov-de
Gennes (BdG) Hamiltonian $\hat{H}_{\text{BdG}}=\frac{1}{2}\sum_{\bm{k}}\hat{\Psi}_{\bm{k}}^{\dagger}H_{\text{BdG}}(\bm{k})\hat{\Psi}_{\bm{k}}$
with 
\begin{align}
H_{{\rm BdG},\Gamma}(\bm{k}) & =\left(\begin{matrix}H_{\bm{k}}^{\frac{3}{2}\oplus\frac{1}{2}}-\mu_{\Gamma} & \Delta_{\Gamma}(\beta_{\Gamma})\\
\Delta_{\Gamma}(\beta_{\Gamma}) & \mu_{\Gamma}-H_{\bm{k}}^{\frac{3}{2}\oplus\frac{1}{2}}
\end{matrix}\right).\label{eq:HBdG}
\end{align}
Here, $\hat{\Psi}_{\bm{k}}^{\dagger}=\left(\hat{C}_{\bm{k}}^{\dagger},\hat{C}_{-\bm{k}}^{T}\gamma\right)$
is a spinor with the creation (destruction) operator $\hat{C}_{\bm{k}}^{\dagger}$
($\hat{C}_{\bm{k}}$) for electrons of the $H_{\bm{k}}^{\frac{3}{2}\oplus\frac{1}{2}}$
and $\gamma=\exp[-i\pi S_{y}^{(3/2)}]\oplus\exp[-i\pi S_{y}^{(1/2)}]$
is the unitary part of the time-reversal operator, and $\mu_{\Gamma}$ is the
Fermi level measured from the four-fold nodal point.

\begin{figure}
\includegraphics{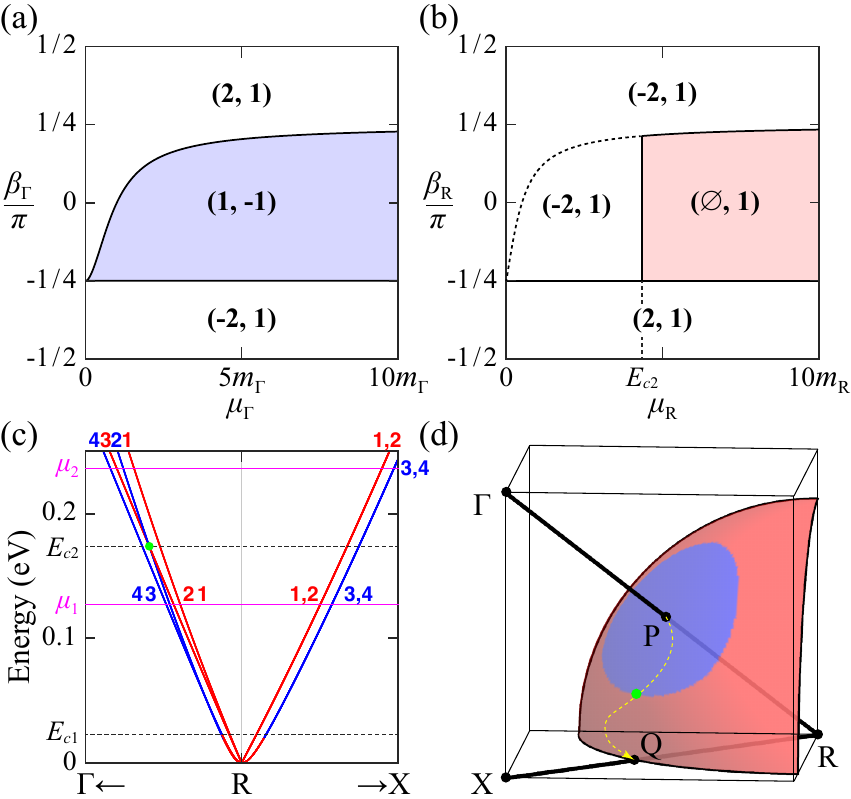} \caption{\label{fig2}(a),(b) Phase diagram of the TSC in RhSi with respect
to $\beta_{\Gamma}$ and $\mu_{\Gamma}$ with $\alpha_{\Gamma}>0$
($\beta_{\text{R}}$ and $\mu_{{\rm R}}$ with $\alpha_{\text{R}}>0$).
A pair of winding numbers $(w_{\text{3D,\ensuremath{\Gamma(\mathrm{R})}}},w_{\text{1D},\Gamma(\mathrm{R})})$
on each region characterizes the topology of the superconducting phase.
$w_{\text{3D},\mathrm{R}}=\varnothing$ indicates the TSC with NRs
in the bulk BdG spectrum. The dashed line represents $\mu_{\text{R}}=E_{c1}(\beta_{\text{R}})$
on which the superconducting gap closes/reopens without changing $w_{3\text{D}}$
and $w_{1\text{D}}$. (c) The band structure around ${\rm R}$ above
the six-fold nodal point colored in red (blue) for positive (negative)
$\delta_{\text{R},n}$. The four bands are labeled as $n=$1, 2, 3,
4 in the order of closeness to ${\rm R}$. (d) The FSS formed by band
2 at $\mu=\mu_{2}$, colored in the same manner as (c).}
\end{figure}

In the weak pairing limit with $\alpha_{\Gamma}$ much smaller than
the energy separation between the bands near the Fermi level, the
BdG spectrum and its topology are largely determined by the intraband
gap functions given by 
\begin{equation}
\begin{split}\delta_{\Gamma,\text{in}}\equiv & \langle\Gamma,\text{in},\bm{k}|\Delta_{\Gamma}(\beta_{\Gamma})|\Gamma,\text{in},\bm{k}\rangle\Bigr|_{k=k_{\text{F,in}}}\\
= & \alpha_{\Gamma}\cos\beta_{\Gamma}(\tan\beta_{\Gamma}+1),\\
\delta_{\Gamma,\text{out}}\equiv & \langle\Gamma,\text{out},\bm{k}|\Delta_{\Gamma}(\beta_{\Gamma})|\Gamma,\text{out},\bm{k}|\rangle\Bigr|_{k=k_{\text{F,out}}}\\
\approx & \alpha_{\Gamma}\cos\beta_{\Gamma}\left[\tan\beta_{\Gamma}-g_{\Gamma}(\mu_{\Gamma}/m_{\Gamma})\right],
\end{split}
\label{eq:proj_gap_Gamma}
\end{equation}
with $g_{\Gamma}(x)=\frac{3x^{2}-2x-1}{3x^{2}+2x+1}$, while the interband
gap functions can be treated perturbatively. The approximation in
Eq.~(\ref{eq:proj_gap_Gamma}) is valid as long as $\alpha_{\Gamma}$
is sufficiently small compared to $\left|3\mu_{\Gamma}+m_{\Gamma}\right|$.

The key insight here is that $\mu_{\Gamma}$ determines the sign of
$\delta_{\Gamma,\text{out}}$ when $|\beta_{\Gamma}|<\pi/4$. This
is most apparent in the case of $\beta_{\Gamma}=0$ in Eq. \eqref{eq:Delta_parameterization}
from the wave function character in Eq.~\eqref{eq:H_Gamma_eigenstate}.
Because $\Delta_{\Gamma}(0)\propto\Delta_{\pm}^{\Gamma}$ in Eq. \eqref{eq:Delta_parameterization},
$\delta_{\Gamma,\text{out}}>0$ for small $\mu_{\Gamma}>0$ since
the outer FSS is mainly composed of the RSW fermion. As $\mu_{\Gamma}$
is increased, however, so does the composition of the KW fermion in
the outer FSS, and $\Delta_{\pm}^{\Gamma}$ gives $\delta_{\Gamma,\text{out}}<0$.
Hence, the variation in the wave function character switches the
sign of $\delta_{\Gamma,\text{out}}$.

In the same manner, we define $\delta_{\text{R,in}}$ and $\delta_{\text{R,out}}$,
which have the same form as those in Eq.~\eqref{eq:proj_gap_Gamma}
but with $g_{{\rm R}}(x)=\frac{2x-1}{2x+1}$, and find that the sign
of $\delta_{\text{R,out}}$ also can be flipped by adjusting $\mu_{\text{R}}>0$
when $|\beta_{\text{R}}|<\pi/4$.

Phase diagrams in Figs. \ref{fig2}(a) and \ref{fig2}(b) summarize
the topological phases at $\Gamma$ and $\text{R}$, respectively. When the whole
BdG spectrum is gapped, a winding number $w_{\text{3D}}$ determines
the topology and the number of helical MZMs on the SBZ \citep{Chiu2016}.
However, as explained below, a TSC with nodal rings (NRs) around
${\rm R}$ can appear for large $\mu_{\text{R}}$. In this case, the
presence/absence of a MZM at $\bar{\Gamma}$ can be determined by
a winding number $w_{\text{1D}}$ \citep{Schnyder2012}. Hence, we
use a pair of winding numbers $(w_{\text{3D}},w_{\text{1D}})$ to
characterize the topology of each region in the phase diagrams.

In the weak pairing limit, $w_{\text{3D}}$ is expressed as \citep{Qi2010}
\begin{align}
w_{{\rm 3D}} & =\frac{1}{2}\sum_{n}{\rm sign}(\delta_{n}){\rm Ch}_{n},
\end{align}
where ${\rm Ch}_{n}$ is the Chern number of the $n$th FSS. We break
$w_{\text{3D}}$ into $w_{\text{3D},\Gamma}$ and $w_{\text{3D},\text{R}}$
according to the centers of FSSs. For $w_{\text{3D},\Gamma}$, we
expect it to be $\pm2$ or $+1$, and the latter occurs only when
$\delta_{\Gamma,\text{out}}\delta_{\Gamma,\text{in}}<0$. Meanwhile,
$w_{\text{3D},\text{R}}=-2\text{sign}(\delta_{\text{R,in}})$ is always
even. Consequently, $w_{{\rm 3D}}$ is odd whenever $\delta_{\Gamma,\text{out}}\delta_{\Gamma,\text{in}}<0$,
and we expect at least one topologically protected MZM at $\bar{\Gamma}$
on SBZ.

Unlike $\delta_{\Gamma,\text{out}}$, the sign flip of $\delta_{\text{R,out}}$
with $\mu_{\text{R}}$ has no effect on $w_{{\rm 3D}}$ since ${\rm Ch}_{{\rm R,out}}=0$.
Nevertheless, it can lead to a nodal TSC in RhSi for sufficiently
large $\mu_{{\rm R}}$, around which the band structure can be understood
by considering the anisotropic $\bm{k}$-quadratic correction to $H_{\boldsymbol{k}}^{1\oplus0}\oplus H_{\boldsymbol{k}}^{1\oplus0}$.
Breaking the artificial isotropy of $H_{\boldsymbol{k}}^{1\oplus0}\oplus H_{\boldsymbol{k}}^{1\oplus0}$
in $\boldsymbol{k}$, it modifies the band structure in Fig.~\ref{fig1}(c)
in two ways. First, there appear band crossings along the ${\rm R}\Gamma$
lines because of the symmetry-enforced band connectivity~\citep{Pshenay-Severin2018}.
Second, the double degeneracy of the model in Eq.~\eqref{eq:H_R}
is lifted in generic momenta in BZ except on $k_{x,y,z}=\pi$ planes
due to the local Kramers theorem from the two-fold screw symmetries
and the time-reversal symmetry~\citep{Chang2017}.

Both effects are clearly shown in Fig.~\ref{fig2}(c). The bands
crossing the Fermi level $\mu_{\text{R}}>0$ are labeled by $n=1,2,3,4$
in the order of closeness to ${\rm R}$, and the color represents
the sign of $\delta_{\text{R},n}$ for $\beta_{\text{R}}=0$ in Eq
\eqref{eq:Delta_parameterization}. As explained, the signs of $\delta_{\text{R},3}$
and $\delta_{\text{R},4}$ are flipped as $\mu_{\text{R}}$ is raised
over $E_{c1}(\beta_{\text{R}})=\frac{m_{\text{R}}}{2}\left(\frac{1+\tan\beta_{\text{R}}}{1-\tan\beta_{\text{R}}}\right)$
at which $g_{\text{R}}(\mu_{\text{R}}/m_{\text{R}})=\tan\beta_{\text{R}}$.
For $\beta_{\text{R}}=0$, $E_{c1}=m_{\text{R}}/2$. The switched
signs are maintained for the higher energies.

The BdG spectrum around ${\rm R}$ remains gapped until $\mu_{\text{R}}$
reaches the band crossing energy $E_{c2}$. For $\mu_{\text{R}}>E_{c2}$,
however, topological NRs in the BdG spectrum appear because the interchange
of wave functions between band 2 and 3 at the band crossing on the
${\rm R}\Gamma$ line makes the signs of $\delta_{\text{R},2(3)}$
on the ${\rm R}\Gamma$ line and on the $k_{x,y,z}=\pi$ planes opposite.
Suppose P (Q) to be a point where the FSS of band 2 and the ${\rm R}\Gamma$
line (${\rm RX}$ line) meet as shown in Fig.~\ref{fig2}(d). Since
$\delta_{{\rm P}}\delta_{{\rm Q}}<0$, there exists at least a point
where the gap function becomes zero for any path between P and Q on
the FSS. As a result, a ring with $\delta_{\text{R},2}=0$ wrapping
the ${\rm R}\Gamma$ line appears on the FSS of band 2. For the same
reason, another NR is found on the FSS of band 3. The stability of
NRs is guaranteed by the 1D winding number for AIII class evaluated
along a loop enclosing the ring~\citep{Schnyder2011}. The rings
from band 2 and 3 have opposite winding numbers of unit magnitude
(see Sec. IV of SM \citep{SM}).

Even in the presence of NRs around ${\rm R}$, a MZM at $\bar{\Gamma}$
is still protected by $w_{{\rm 1D}}$ evaluated along a TRI loop ${\cal L}$
connecting $\Gamma$ with another TRIM that is projected to $\bar{\Gamma}$
on the SBZ without intersecting with NRs in the BZ~\citep{Schnyder2011}.
In the weak pairing limit, $w_{{\rm 1D}}$ can be easily evaluated
from \citep{Qi2010} 
\begin{align}
w_{{\rm 1D}}= & \prod_{n\in\text{FSS}'}\text{sign}(\delta_{n}),
\end{align}
where FSS$'$ denotes the FSSs which intersect with ${\cal L}$. Note
that the four FSSs near ${\rm R}$ do not contribute to $w_{{\rm 1D}}$
due to the local Kramers degeneracies implying $\delta_{\text{R},2n-1}=\delta_{\text{R},2n}$
for $n=1,2$. Thus, a MZM at $\bar{\Gamma}$ is expected for any surfaces
of the crystal as long as $\delta_{\Gamma,{\rm in}}\delta_{\Gamma,{\rm out}}<0$,
which is consistent with the condition in the fully gapped phase.

\textit{TSC phase diagram in various interactions.}--- To investigate
what kind of interactions realize the $s_{+}\oplus s_{-}$ gap functions,
we consider three types of electron-electron interactions: 
\begin{equation}
\hat{H}_{\text{int}}=\frac{U}{2}\sum_{i}\hat{\rho}_{i}\hat{\rho}_{i}+\frac{V}{2}\sum_{\langle i,j\rangle}\hat{\rho}_{i}\hat{\rho}_{j}+\frac{J}{2}\sum_{\langle i,j\rangle}\bm{\hat{S}}_{i}\cdot\bm{\hat{S}}_{j}.\label{eq:H_int}
\end{equation}
Here, $(\hat{\rho}_{i},\hat{\boldsymbol{S}}_{i})=(\hat{C}_{i,\uparrow}^{\dagger},\hat{C}_{i,\downarrow}^{\dagger})(s_{0},\boldsymbol{s})(\hat{C}_{i,\uparrow},\hat{C}_{i,\downarrow})^{T}$
with the spin Pauli matrices $\bm{s}$. $U$, $V$ and $J$ represent
the renormalized on-site Coulomb interaction, the nearest-neighbor
Coulomb interaction, and the nearest-neighbor (anti)ferromagnetic
exchange interaction, respectively~\citep{Vafek2017}. Considering
only the time-reversal invariant pairing channels from $\hat{H}_{\text{int}}$
respecting spatial symmetries of RhSi, the pairing interaction $\hat{H}_{\text{pair}}$
can be expressed as 
\begin{eqnarray}
\hat{H}_{\text{pair}} & = & \sum_{i=0}^{8}\frac{U_{i}}{4}\sum_{\bm{k},\bm{p}}\hat{\Pi}_{i}(\bm{k})\hat{\Pi}_{i}^{\dagger}(\bm{p}),\label{eq:Pairing interaction}
\end{eqnarray}
where $\hat{\Pi}_{i}(\bm{k})=\hat{C}_{-\boldsymbol{k}}^{T}\gamma M_{i,\bm{k}}^{\dagger}\hat{C}_{\boldsymbol{k}}$,
and $U_{i}$'s are the coupling constants of the pairing channels
$\hat{\Pi}_{i}\hat{\Pi}_{i}^{\dagger}$ with $U_{0}=U/4$, $U_{1,5}=V-3J$,
and $U_{2,3,4,6,7,8}=V+J$. Here, $M_{i,\bm{k}}$ are $\bm{k}$-dependent
matrices characterizing the pairing channels (see Sec. I of SM \citep{SM}).

\begin{figure}[t]
\includegraphics{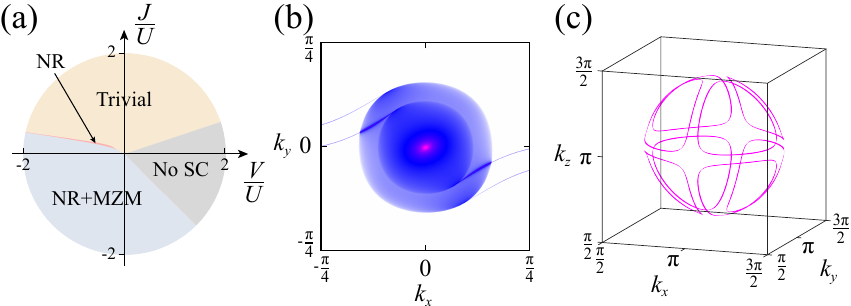} \caption{\label{fig3}(a) Phase diagram in $(V/U,J/U)$ at $\mu=1.3m_{\Gamma}$
with $U>0$. ``No SC'' and ``Trivial'' mean no superconductivity
and the topologically trivial superconducting phase, respectively,
whereas ``NR'' and ``MZM'' represent the superconducting phases
with the NRs around ${\rm R}$, and a MZM at $\bar{\Gamma}$, respectively.
(b) Surface spectral weight around $\bar{\Gamma}$ at $E=0$. (c)
Nodal rings around ${\rm R}$ in the BZ.}
\end{figure}

Solving the linearized gap equation with $\hat{H}_{\text{pair}}$,
the gap function $\Delta(\bm{k})=\sum_{i=0}^{8}\Delta_{i}M_{i,\bm{k}}$
with the highest transition temperature is obtained. The relationship
between $\Delta(\boldsymbol{k})$ and Eq.~(\ref{eq:Delta_parameterization})
can be understood by projecting $\Delta(\boldsymbol{k})$ to the bases
of $H_{\boldsymbol{k}}^{\frac{3}{2}\oplus\frac{1}{2}}$ and $H_{\boldsymbol{k}}^{1\oplus0}$
at $\Gamma$ and ${\rm R}$, respectively. At $\Gamma$, $\Delta(\boldsymbol{k})\sim(\Delta_{0}-\Delta_{1})I_{6}+(\Delta_{2}-\Delta_{3})\Delta_{\pm}^{\Gamma}$,
whereas at ${\rm R}$, $\Delta(\boldsymbol{k})\sim\Delta_{0}I_{4}+\Delta_{4}\Delta_{\pm}^{\text{R}}$
up to the zeroth order in $\boldsymbol{k}$. For $U<0$, the trivial
gap function largely prevails. Hence, we focus on $U>0$.

Figure~\ref{fig3}(a) shows the phase diagram in $(V/U,J/U)$ for
$U>0$ at $\mu=\mu_{\Gamma}=\mu_{\text{R}}-0.43\;{\rm eV}=1.3m_{\Gamma}$.
In the blue ``NR+MZM'' region, we find that the $s_{+}\oplus s_{-}$
gap functions are dominant and the resultant TSC exhibits a
MZM at $\bar{\Gamma}$ and the NRs near ${\rm R}$ as shown in Figs.~\ref{fig3}(b)
and \ref{fig3}(c), respectively. This region is approximately covered
by $V<-J$ with $J<0$, where the ferromagnetic interaction is stronger
than the nearest-neighborhood Coulomb interaction.
In a narrow region in red at the boundary of the blue region, the TSC with only
the NRs from $\Delta_{\pm}^{\text{R}}$ appears. The beige region
marked by ``Trivial'' approximately overlaps with $V<3J$ with $J>0$,
where the topologically trivial SC arises from the spin-singlet
pairings due to the anti-ferromagnetic interaction $J>0$ in contrast
to the ``NR+MZM'' region. Near the phase boundaries, TSC phases
with additional NRs around $\Gamma$ or R may appear due to a complicated
combination of various gap functions (see Sec. V of SM \citep{SM}).

\textit{Discussion.}--- The $s_{+}\oplus s_{-}$ gap functions in
our work should be distinguished from the momentum-dependent $s_{\pm}$
gap functions usually discussed in iron-based superconductors~\citep{Chubukov2012}.
We find that an analogous gap function for our system, corresponding
to $\alpha_{\Gamma}\alpha_{\text{R}}<0$ in Eq. \eqref{eq:Delta_parameterization}
\citep{Huang2020,Gao2020}, is not favored by the electron-electron
interaction of Eq.~\eqref{eq:H_int} (see Sec. V of SM \citep{SM}).

Regarding the experimental realization, we want to refer to the report on the multi-gap superconductivity
in RhGe~\citep{Tsvyshchenko2016} whose electronic structure is akin to RhSi.
Considering weak ferromagnetism in RhGe,  it is expected that the TSC arising from the $s_{+}\oplus s_{-}$ gap functions
could be realized by varying its chemical compositions~\citep{Salamatin2021}.
Also, we expect that our result could be applied to the multi-gap superconductor BeAu~\citep{Matthias1959,Rebar2019, Khasanov2020} in B20 structure as well as other noncentrosymmetric superconductors such as Li$_{2}$X$_{3}$B (X$=$Pd, Pt) ~\citep{Gao2020}, PtSbS~\citep{Mizutani2019}, and BaPtP~\citep{Okamoto2020} supporting coupled multifold fermions.

The TSC from the $s_{+}\oplus s_{-}$
gap functions can be investigated by probing the bulk and surface
properties, respectively. To discern the gap function from the trivial
$s$-wave gap function, magnetic resonance techniques and angle-resolved
photoemission spectroscopy can be used to identify the spin state
of the gap function and the size of the superconducting gap on each
FSS, respectively \citep{Ishida1998,Mou2015}. Regarding the
response from the surface, drastic anisotropy in response to magnetic fields could be a smoking gun of the MZMs on the surface \citep{Chung2009,Chung2013,Chirolli2019}.




\begin{acknowledgments}
This work was supported by the National Research Foundation of Korea
(NRF) grant funded by the Korea government (MSIT) (No. 2018R1A2B6007837)
and Creative-Pioneering Researchers Program through Seoul National
University (SNU). C. Y. was supported by NRF of Korea under Grant
No. 2016H1A2A1907780 through the Global PhD Fellowship Program. S.B.C.
was supported by NRF grants funded by the Korea government (MSIT)
(No. 2020R1A2C1007554) and the Ministry of Education (2018R1A6A1A06024977). 
\end{acknowledgments}

\bibliography{ref}

\end{document}


\widetext \makeatletter \pdfstringdefDisableCommands{\let\bm\@firstofone}
\pdfstringdefDisableCommands{\let\bm\@firstofone} 
\makeatother
\setcounter{figure}{0} 
\global\long\def\thefigure{S\arabic{figure}}%

\begin{center}
\textbf{\large{}Supplemental Material:}\\
\textbf{\large{}Topological Multiband $s$-wave Superconductivity
in Coupled Multifold Fermions}{\large\par}
\par\end{center}

\hypersetup{linkcolor=black}\tableofcontents{}
\begin{center}
\vfill{}
\par\end{center}

\pagebreak{}

\hypersetup{linkcolor=magenta}

\section{Lattice and Continuum Models for RhSi}

\subsection{Tight-binding model\label{sec:TB model for RhSi}}

In the main text, we use the tight-binding model proposed in Ref.~\citep{Chang2017}.
The model uses a $s$-wave-like orbital at Wyckoff positions with
the minimum multiplicity of 4, and the corresponding four sites $A=(0,0,0)$,
$B=(\frac{1}{2},\frac{1}{2},0)$, $C=(\frac{1}{2},0,\frac{1}{2})$
and $D=(0,\frac{1}{2},\frac{1}{2})$ are included in the unit cell.
(Here we set the lattice constant to unity.) The resultant normal
phase Hamiltonian in the basis $\left(A_{\uparrow},A_{\downarrow},C_{\uparrow},C_{\downarrow},B_{\uparrow},B_{\downarrow},D_{\uparrow},D_{\downarrow}\right)$
is given by
\begin{align}
H_{\text{TB}}(\bm{k})= & H_{\text{TB}}^{\text{w/o SOC}}(\bm{k})+v_{r,1}M_{2,\bm{k}}+v_{r,2}M_{3,\bm{k}}+v_{r,3}M_{4,\bm{k}}+v_{s,1}M_{6,\bm{k}}+v_{s,2}M_{7,\bm{k}}+v_{s,3}M_{8,\bm{k}},\label{eq:lattice_tight-binding_model}\\
H_{\text{TB}}^{\text{w/o SOC}}(\bm{k})= & v_{2}\bigg[\sum_{i=x,y,z}\cos k_{i}-\varepsilon_{\text{offset}}\bigg]M_{0}+v_{1}M_{1,\bm{k}}+v_{p}M_{5,\bm{k}},
\end{align}
where $\varepsilon_{\text{offset}}=-v_{1}+3v_{2}+v_{r,1}-v_{r,2}$
is introduced to set the energy level of the RSW nodal point at $\Gamma$
as zero. The nine $M_{i,\bm{k}}$ matrices are given by 
\begin{subequations}
\begin{align}
M_{0}= & I_{8},\\
M_{1,\bm{k}}= & \tau_{1}\cos\frac{k_{x}}{2}\cos\frac{k_{y}}{2}+\tau_{1}\mu_{1}\cos\frac{k_{y}}{2}\cos\frac{k_{z}}{2}+\mu_{1}\cos\frac{k_{z}}{2}\cos\frac{k_{x}}{2},\\
M_{2,\bm{k}}= & \tau_{2}\mu_{3}s_{2}\cos\frac{k_{x}}{2}\cos\frac{k_{y}}{2}+\tau_{2}\mu_{1}s_{3}\cos\frac{k_{y}}{2}\cos\frac{k_{z}}{2}+\mu_{2}s_{1}\cos\frac{k_{z}}{2}\cos\frac{k_{x}}{2},\\
M_{3,\bm{k}}= & \tau_{2}s_{3}\cos\frac{k_{x}}{2}\cos\frac{k_{y}}{2}+\tau_{1}\mu_{2}s_{1}\cos\frac{k_{y}}{2}\cos\frac{k_{z}}{2}+\tau_{3}\mu_{2}s_{2}\cos\frac{k_{z}}{2}\cos\frac{k_{x}}{2},\\
M_{4,\bm{k}}= & \tau_{2}\mu_{3}s_{1}\sin\frac{k_{x}}{2}\sin\frac{k_{y}}{2}+\tau_{2}\mu_{1}s_{2}\sin\frac{k_{y}}{2}\sin\frac{k_{z}}{2}+\mu_{2}s_{3}\sin\frac{k_{z}}{2}\sin\frac{k_{x}}{2},\label{eq:Miks}\\
M_{5,\bm{k}}= & \tau_{2}\mu_{3}\cos\frac{k_{x}}{2}\sin\frac{k_{y}}{2}+\tau_{2}\mu_{1}\cos\frac{k_{y}}{2}\sin\frac{k_{z}}{2}+\mu_{2}\cos\frac{k_{z}}{2}\sin\frac{k_{x}}{2},\\
M_{6,\bm{k}}= & \tau_{1}s_{1}\sin\frac{k_{x}}{2}\cos\frac{k_{y}}{2}+\tau_{1}\mu_{1}s_{2}\sin\frac{k_{y}}{2}\cos\frac{k_{z}}{2}+\mu_{1}s_{3}\sin\frac{k_{z}}{2}\cos\frac{k_{x}}{2},\\
M_{7,\bm{k}}= & \tau_{1}s_{2}\cos\frac{k_{x}}{2}\sin\frac{k_{y}}{2}+\tau_{1}\mu_{1}s_{3}\cos\frac{k_{y}}{2}\sin\frac{k_{z}}{2}+\mu_{1}s_{1}\cos\frac{k_{z}}{2}\sin\frac{k_{x}}{2},\\
M_{8,\bm{k}}= & \tau_{1}\mu_{3}s_{3}\cos\frac{k_{x}}{2}\sin\frac{k_{y}}{2}-\tau_{2}\mu_{2}s_{1}\cos\frac{k_{y}}{2}\sin\frac{k_{z}}{2}+\tau_{3}\mu_{1}s_{2}\cos\frac{k_{z}}{2}\sin\frac{k_{x}}{2},
\end{align}
\end{subequations}
 where $\tau_{i}$ and $\mu_{i}$ are Pauli matrices related to the
orbital degrees of freedom at the four Wyckoff positions $A$, $B$,
$C$ and $D$, and $s_{i}$ are Pauli matrices operating on the real
spin degrees of freedom. $\varepsilon_{\text{offset}}$ is fitted
to $-0.04$ eV. In the main text, we take $v_{1}=0.55$, $v_{2}=0.16$,
$v_{p}=-0.76$, $v_{r,1}=0,$ $v_{r,2}=-0.03$, $v_{r,3}=0.01$, $v_{s,1}=0$,
$v_{s,2}=0$, and $v_{s,3}=0$ in eV which are different from Ref.
\citep{Chang2017} where $v_{s,1}=-0.04$ is used. Considering that
$M_{6-8,\boldsymbol{k}}$ are odd functions of $\boldsymbol{k}$ around
$\Gamma$ and R, the role of small $v_{s,i}$'s is to slightly modify
the Fermi velocity of the bands which is mainly determined by $v_{p}$.
Also, because of the limited usage of atomic basis, the tight-binding
model given here does not accurately capture the electronic structure
all over the range we are interested in. For these reasons, there
is an unavoidable ambiguity in the numerical values for $v_{s,i}$'s.
However, we find that the result we report remains qualitatively unchanged
in a reasonable range of $v_{s,i}$'s, as explained in detail in Sec.
\ref{sec:V}. Hence, we take $v_{s,i}=0$ for simplicity.

\subsection{Continuum model near the $\Gamma$ point\label{sec:continuum_model_G}}

At $\Gamma$, $H_{\text{TB}}^{\text{w/o SOC}}(\bm{k})$ in Eq.~(\ref{eq:lattice_tight-binding_model})
yields 6-fold and 2-fold nodal points at the energy level $\varepsilon_{\text{6-fold}}$
and $\varepsilon_{\text{2-fold}}$, respectively. The Fermi energy
lies near the 6-fold nodal point for RhSi. As\textbf{ }the energy
difference $\varepsilon_{\text{2-fold}}-\varepsilon_{\text{6-fold}}$
is relatively huge, approximately $2$ eV in RhSi, we can obtain the
effective $6\times6$ Hamiltonian near $\Gamma$ by projecting out
the degrees of freedom of the 2-fold nodal point:
\begin{equation}
H^{(\Gamma)}(\bm{k})=\mathbb{P}_{\Gamma}^{\dagger}H_{\text{TB}}(\bm{k}){\cal \mathbb{P}}_{\Gamma},
\end{equation}
where $\mathbb{P}_{\Gamma}$ is a $8\times6$ matrix given by
\begin{equation}
\mathbb{P}_{\Gamma}=\frac{1}{2}\left(\begin{matrix}-z_{1}^{*} & \frac{-2i}{\sqrt{6}} & \frac{z_{1}}{\sqrt{3}} & 0 & \frac{i}{\sqrt{3}} & \frac{2z_{1}}{\sqrt{6}}\\
0 & \frac{-z_{1}^{*}}{\sqrt{3}} & \frac{-2i}{\sqrt{6}} & -z_{1} & \frac{-2z_{1}^{*}}{\sqrt{6}} & \frac{-i}{\sqrt{3}}\\
z_{1}^{*} & \frac{-2i}{\sqrt{6}} & \frac{z_{1}}{\sqrt{3}} & 0 & \frac{i}{\sqrt{3}} & \frac{-2z_{1}}{\sqrt{6}}\\
0 & \frac{z_{1}^{*}}{\sqrt{3}} & \frac{-2i}{\sqrt{6}} & z_{1} & \frac{2z_{1}^{*}}{\sqrt{6}} & \frac{-i}{\sqrt{3}}\\
z_{1} & \frac{2i}{\sqrt{6}} & \frac{z_{1}^{*}}{\sqrt{3}} & 0 & \frac{-i}{\sqrt{3}} & \frac{-2z_{1}^{*}}{\sqrt{6}}\\
0 & \frac{z_{1}}{\sqrt{3}} & \frac{2i}{\sqrt{6}} & z_{1}^{*} & \frac{2z_{1}}{\sqrt{6}} & \frac{i}{\sqrt{3}}\\
-z_{1} & \frac{2i}{\sqrt{6}} & \frac{-z_{1}^{*}}{\sqrt{3}} & 0 & \frac{-i}{\sqrt{3}} & \frac{2z_{1}^{*}}{\sqrt{6}}\\
0 & \frac{-z_{1}}{\sqrt{3}} & \frac{2i}{\sqrt{6}} & -z_{1}^{*} & \frac{-2z_{1}}{\sqrt{6}} & \frac{i}{\sqrt{3}}
\end{matrix}\right)
\end{equation}
with $z_{1}=e^{\frac{\pi i}{4}}$. ${\cal \mathbb{P}}_{\Gamma}$ is
constructed from the eigenvectors of $H_{\text{TB}}^{\text{w/o SOC}}(\boldsymbol{0})$
satisfying $H_{\text{TB}}^{\text{w/o SOC}}(\boldsymbol{0}){\cal \mathbb{P}}_{\Gamma}=\varepsilon_{\text{6-fold}}\mathbb{P}_{\Gamma}$.
Expanding $H^{(\Gamma)}(\bm{k})$ to linear order in $\bm{k}$, we
have
\begin{align}
H^{(\Gamma)}(\bm{k})= & H_{\bm{k}}^{\frac{3}{2}\oplus\frac{1}{2}}+\sum_{i=1,2,3}v_{s,i}\delta H_{s,i}^{(\Gamma)}(\bm{k})+O(k^{2}),\\
H_{\bm{k}}^{\frac{3}{2}\oplus\frac{1}{2}}= & \left(\begin{matrix}v_{\Gamma}\bm{k}\cdot\bm{S}^{(\frac{3}{2})} & T_{\Gamma,\bm{k}}\\
T_{\Gamma,\bm{k}}^{\dagger} & 2v_{\Gamma}\bm{k}\cdot\bm{S}^{(\frac{1}{2})}-m_{\Gamma}
\end{matrix}\right),
\end{align}
where $m_{\Gamma}=3v_{r,1}-3v_{r,2}$, $v_{\Gamma}=v_{p}/3$ and
\begin{equation}
T_{\Gamma,k}^{\dagger}=\frac{v_{\Gamma}}{2\sqrt{2}}\left(\begin{matrix}-\sqrt{3}k_{+} & 2k_{z} & k_{-} & 0\\
0 & -k_{+} & 2k_{z} & \sqrt{3}k_{-}
\end{matrix}\right)
\end{equation}
with $k_{\pm}=k_{x}\pm ik_{y}$. Here, $\delta H_{s,i}^{(\Gamma)}$
is the matrix $M_{5+i,\boldsymbol{k}}$ expanded to the linear order
in $\boldsymbol{k}$:
\begin{align}
\delta H_{s,1}^{(\Gamma)}(\boldsymbol{k})= & \left(\begin{array}{cccccc}
-\frac{k_{z}}{2} & 0 & 0 & -\frac{k_{x}+ik_{y}}{2} & 0 & 0\\
0 & \frac{k_{z}}{2} & -\frac{k_{x}-ik_{y}}{2} & 0 & 0 & 0\\
0 & -\frac{k_{x}+ik_{y}}{2} & -\frac{k_{z}}{2} & 0 & 0 & 0\\
-\frac{k_{x}-ik_{y}}{2} & 0 & 0 & \frac{k_{z}}{2} & 0 & 0\\
0 & 0 & 0 & 0 & \frac{k_{z}}{2} & \frac{k_{x}-ik_{y}}{2}\\
0 & 0 & 0 & 0 & \frac{k_{x}+ik_{y}}{2} & -\frac{k_{z}}{2}
\end{array}\right),\\
\delta H_{s,2}^{(\Gamma)}(\boldsymbol{k})= & \left(\begin{array}{cccccc}
0 & -\frac{k_{x}}{2\sqrt{3}} & \frac{k_{z}}{2\sqrt{3}} & \frac{ik_{y}}{2} & -\frac{k_{x}}{\sqrt{6}} & -\frac{k_{z}}{\sqrt{6}}\\
-\frac{k_{x}}{2\sqrt{3}} & -\frac{k_{z}}{3} & \frac{2k_{x}+ik_{y}}{6} & -\frac{k_{z}}{2\sqrt{3}} & \frac{k_{z}}{3\sqrt{2}} & \frac{k_{x}+2ik_{y}}{3\sqrt{2}}\\
\frac{k_{z}}{2\sqrt{3}} & \frac{2k_{x}-ik_{y}}{6} & \frac{k_{z}}{3} & -\frac{k_{x}}{2\sqrt{3}} & -\frac{k_{x}-2ik_{y}}{3\sqrt{2}} & \frac{k_{z}}{3\sqrt{2}}\\
-\frac{ik_{y}}{2} & -\frac{k_{z}}{2\sqrt{3}} & -\frac{k_{x}}{2\sqrt{3}} & 0 & -\frac{k_{z}}{\sqrt{6}} & \frac{k_{x}}{\sqrt{6}}\\
-\frac{k_{x}}{\sqrt{6}} & \frac{k_{z}}{3\sqrt{2}} & -\frac{k_{x}+2ik_{y}}{3\sqrt{2}} & -\frac{k_{z}}{\sqrt{6}} & -\frac{k_{z}}{6} & -\frac{k_{x}-ik_{y}}{6}\\
-\frac{k_{z}}{\sqrt{6}} & \frac{k_{x}-2ik_{y}}{3\sqrt{2}} & \frac{k_{z}}{3\sqrt{2}} & \frac{k_{x}}{\sqrt{6}} & -\frac{k_{x}+ik_{y}}{6} & \frac{k_{z}}{6}
\end{array}\right),\\
\delta H_{s,3}^{(\Gamma)}(\boldsymbol{k})= & \left(\begin{array}{cccccc}
0 & -\frac{ik_{y}}{2\sqrt{3}} & \frac{k_{z}}{2\sqrt{3}} & -\frac{k_{x}}{2} & \frac{ik_{y}}{2\sqrt{6}} & \frac{k_{z}}{2\sqrt{6}}\\
\frac{ik_{y}}{2\sqrt{3}} & \frac{k_{z}}{3} & \frac{k_{x}+2ik_{y}}{6} & -\frac{k_{z}}{2\sqrt{3}} & \frac{k_{z}}{6\sqrt{2}} & -\frac{2k_{x}+ik_{y}}{6\sqrt{2}}\\
\frac{k_{z}}{2\sqrt{3}} & \frac{k_{x}-2ik_{y}}{6} & -\frac{k_{z}}{3} & -\frac{ik_{y}}{2\sqrt{3}} & \frac{2k_{x}-ik_{y}}{6\sqrt{2}} & \frac{k_{z}}{6\sqrt{2}}\\
-\frac{k_{x}}{2} & -\frac{k_{z}}{2\sqrt{3}} & \frac{ik_{y}}{2\sqrt{3}} & 0 & \frac{k_{z}}{2\sqrt{6}} & \frac{ik_{y}}{2\sqrt{6}}\\
-\frac{ik_{y}}{2\sqrt{6}} & \frac{k_{z}}{6\sqrt{2}} & \frac{2k_{x}+ik_{y}}{6\sqrt{2}} & \frac{k_{z}}{2\sqrt{6}} & -\frac{k_{z}}{3} & -\frac{k_{x}-ik_{y}}{3}\\
\frac{k_{z}}{2\sqrt{6}} & \frac{ik_{y}-2k_{x}}{6\sqrt{2}} & \frac{k_{z}}{6\sqrt{2}} & -\frac{ik_{y}}{2\sqrt{6}} & -\frac{k_{x}+ik_{y}}{3} & \frac{k_{z}}{3}
\end{array}\right).
\end{align}
As long as $v_{s,i}$'s are much small in magnitude compared to $v_{p}$,
the role of $\delta H_{s,1-3}$ is just to introduce small anisotropy
in the shape of the Fermi surface given by $H^{(\Gamma)}(\bm{k})$.
Thus, for $|v_{s,i}|\ll|v_{p}|$, we can neglect them to a good approximation,
and thus, $H^{(\Gamma)}(\bm{k})$ can be approximated as $H_{\bm{k}}^{\frac{3}{2}\oplus\frac{1}{2}}$
which we use in the main text.

\subsection{Continuum model near the ${\rm R}$ point\label{sec:continuum_model_R}}

\begin{figure}[th]
\includegraphics[width=1\textwidth]{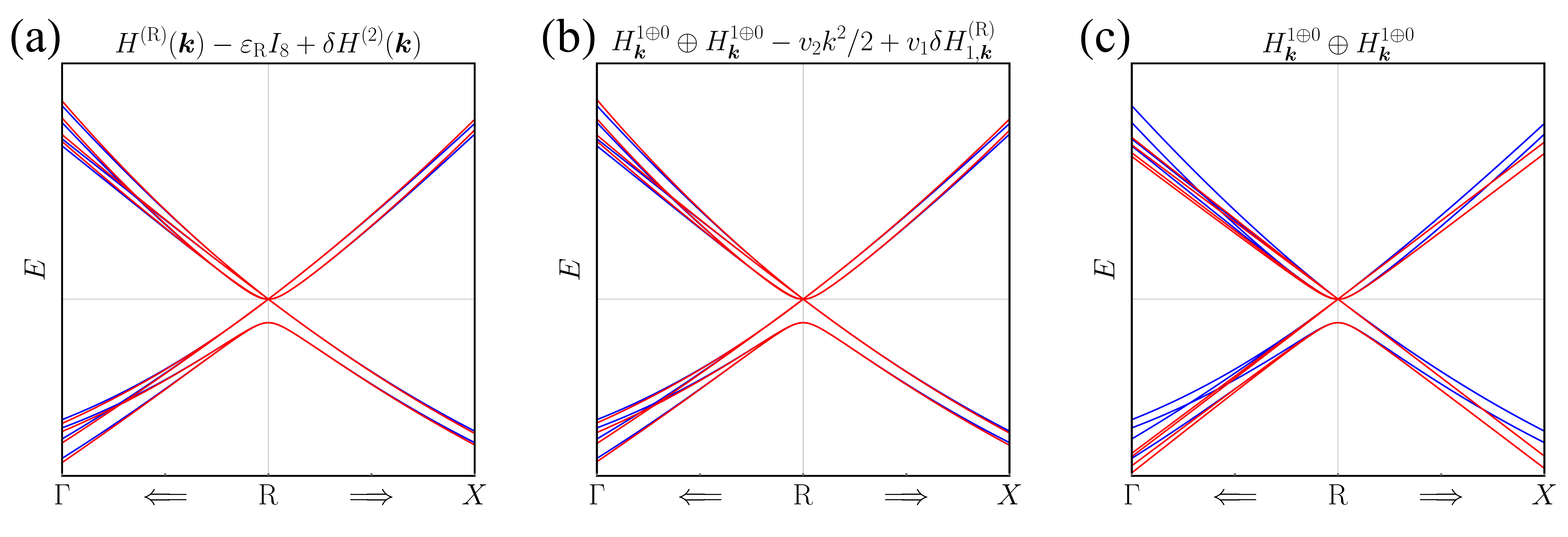}

\caption{\label{fig:figS1}\hspace{0pt}Comparison between the tight-binding
band structure (blue line) and the continuum band structure (red line)
from (a) $H^{(\text{R})}(\boldsymbol{k})-\varepsilon_{\text{R}}+\delta H^{(2)}(\boldsymbol{k})$
including all $\boldsymbol{k}$-quadratic terms, (b) $H_{\bm{k}}^{1\oplus0}\oplus H_{\bm{k}}^{1\oplus0}-\frac{v_{2}k^{2}}{2}+v_{1}\delta H_{1,\boldsymbol{k}}^{(\text{R})}$,
and (c) $H_{\bm{k}}^{1\oplus0}\oplus H_{\bm{k}}^{1\oplus0}$.}
\end{figure}

At ${\rm R}$, 6-fold and 2-fold nodal points are obtained from $H_{\text{TB}}(\boldsymbol{k}_{\text{R}})$
with $\boldsymbol{k}_{\text{R}}=(\pi,\pi,\pi)$. The matrix of eigenvectors
for $H_{\text{TB}}(\boldsymbol{k}_{\text{R}})$ is given by 
\begin{equation}
\mathbb{P}_{\text{R}}=\frac{e^{-i\frac{\pi}{4}}}{2}\left(\begin{array}{cccc|cccc}
1 & \frac{i}{\sqrt{2}} & 0 & \frac{i}{\sqrt{2}} & -1 & \frac{i}{\sqrt{2}} & 0 & \frac{i}{\sqrt{2}}\\
0 & \frac{1}{\sqrt{2}} & i & -\frac{1}{\sqrt{2}} & 0 & -\frac{1}{\sqrt{2}} & i & \frac{1}{\sqrt{2}}\\
i & \frac{1}{\sqrt{2}} & 0 & \frac{1}{\sqrt{2}} & -i & \frac{1}{\sqrt{2}} & 0 & \frac{1}{\sqrt{2}}\\
0 & \frac{i}{\sqrt{2}} & 1 & -\frac{i}{\sqrt{2}} & 0 & -\frac{i}{\sqrt{2}} & 1 & \frac{i}{\sqrt{2}}\\
-1 & \frac{i}{\sqrt{2}} & 0 & \frac{i}{\sqrt{2}} & -1 & -\frac{i}{\sqrt{2}} & 0 & -\frac{i}{\sqrt{2}}\\
0 & -\frac{1}{\sqrt{2}} & i & \frac{1}{\sqrt{2}} & 0 & -\frac{1}{\sqrt{2}} & -i & \frac{1}{\sqrt{2}}\\
-i & \frac{1}{\sqrt{2}} & 0 & \frac{1}{\sqrt{2}} & -i & -\frac{1}{\sqrt{2}} & 0 & -\frac{1}{\sqrt{2}}\\
0 & -\frac{i}{\sqrt{2}} & 1 & \frac{i}{\sqrt{2}} & 0 & -\frac{i}{\sqrt{2}} & -1 & \frac{i}{\sqrt{2}}
\end{array}\right).
\end{equation}
Here, the 1, 2, 3, 5, 6, 7-th columns and the 4, 8-th columns of $\mathbb{P}_{\text{R}}$
represent the eigenvectors for the 6-fold and 2-fold nodal points,
respectively. In contrast to the case at $\Gamma$, these nodal points
are not energetically distant, and hence, the whole 8 degrees of freedom
should be considered on equal footing without projecting out some
of them. The effective Hamiltonian for electrons near ${\rm R}$ is
obtained by expanding $H_{\text{TB}}(\boldsymbol{k}_{\text{R}}+\bm{k})$
as 
\begin{equation}
H^{(\text{R})}(\bm{k})=\mathbb{P}_{\text{R}}^{\dagger}H_{\text{TB}}(\boldsymbol{k}_{\text{R}}+\bm{k}){\cal \mathbb{P}}_{\text{R}}.
\end{equation}
To linear order in $\bm{k}$, we have 
\begin{align}
H^{(\text{R})}(\bm{k})= & \varepsilon_{\text{R}}I_{8}+H_{\boldsymbol{k}}^{1\oplus0}\oplus H_{\boldsymbol{k}}^{1\oplus0}+\sum_{i=1,2,3}v_{s,i}\delta H_{s,i,\boldsymbol{k}}^{(\text{R})}+O(k^{2}),\label{eq:HR}\\
H_{\boldsymbol{k}}^{1\oplus0}= & \left(\begin{matrix}-v_{\text{R}}\bm{k}\cdot\bm{S}^{(1)} & T_{\text{R},\bm{k}}\\
T_{\text{R},\bm{k}}^{\dagger} & -m_{\text{R}}
\end{matrix}\right),
\end{align}
where $\varepsilon_{\text{R}}=3v_{2}+v_{r,3}-\varepsilon_{\text{offset}}$,
$v_{\text{R}}=-v_{p}/2$, $m_{\text{R}}=-4v_{r,3}$ and 
\begin{equation}
T_{\text{R},k}^{\dagger}=v_{\text{R}}\left(\begin{matrix}\frac{k_{+}}{\sqrt{2}} & -k_{z} & \frac{-k_{-}}{\sqrt{2}}\end{matrix}\right).\label{eq:TR}
\end{equation}
$\delta H_{s,i}^{(\text{R})}$ is the matrix $M_{5+i,\boldsymbol{k}_{\text{R}}+\boldsymbol{k}}$
expanded to the linear order in $\boldsymbol{k}$: 
\begin{align}
\delta H_{s,1}^{(\text{R})}(\boldsymbol{k})= & \frac{1}{2\sqrt{2}}\left(\begin{matrix}0 & k_{y} & 0 & -k_{y} & 0 & -k_{x} & -i\sqrt{2}k_{z} & -k_{x}\\
k_{y} & 0 & -k_{y} & 0 & k_{x} & 0 & k_{x} & i\sqrt{2}k_{z}\\
0 & -k_{y} & 0 & -k_{y} & i\sqrt{2}k_{z} & -k_{x} & 0 & k_{x}\\
-k_{y} & 0 & -k_{y} & 0 & k_{x} & -i\sqrt{2}k_{z} & -k_{x} & 0\\
0 & k_{x} & -i\sqrt{2}k_{z} & k_{x} & 0 & -k_{y} & 0 & k_{y}\\
-k_{x} & 0 & -k_{x} & i\sqrt{2}k_{z} & -k_{y} & 0 & k_{y} & 0\\
i\sqrt{2}k_{z} & k_{x} & 0 & -k_{x} & 0 & k_{y} & 0 & k_{y}\\
-k_{x} & -i\sqrt{2}k_{z} & k_{x} & 0 & k_{y} & 0 & k_{y} & 0
\end{matrix}\right),\\
\delta H_{s,2}^{(\text{R})}(\boldsymbol{k})= & \frac{1}{2\sqrt{2}}\left(\begin{array}{cccccccc}
0 & -ik_{x} & 0 & ik_{x} & 0 & k_{y} & -\sqrt{2}k_{z} & k_{y}\\
ik_{x} & 0 & ik_{x} & 0 & k_{y} & 0 & -k_{y} & -\sqrt{2}k_{z}\\
0 & -ik_{x} & 0 & -ik_{x} & \sqrt{2}k_{z} & -k_{y} & 0 & k_{y}\\
-ik_{x} & 0 & ik_{x} & 0 & k_{y} & \sqrt{2}k_{z} & k_{y} & 0\\
0 & k_{y} & -\sqrt{2}k_{z} & k_{y} & 0 & ik_{x} & 0 & -ik_{x}\\
k_{y} & 0 & -k_{y} & -\sqrt{2}k_{z} & -ik_{x} & 0 & -ik_{x} & 0\\
\sqrt{2}k_{z} & -k_{y} & 0 & k_{y} & 0 & ik_{x} & 0 & ik_{x}\\
k_{y} & \sqrt{2}k_{z} & k_{y} & 0 & ik_{x} & 0 & -ik_{x} & 0
\end{array}\right),\\
\delta H_{s,3}^{(\text{R})}(\boldsymbol{k})= & \frac{i}{2\sqrt{2}}\left(\begin{matrix}0 & 0 & \sqrt{2}k_{z} & 0 & 0 & k_{x}+k_{y} & 0 & k_{x}-k_{y}\\
0 & 0 & 0 & -\sqrt{2}k_{z} & -k_{x}+k_{y} & 0 & -k_{x}-k_{y} & 0\\
-\sqrt{2}k_{z} & 0 & 0 & 0 & 0 & k_{x}-k_{y} & 0 & -k_{x}-k_{y}\\
0 & \sqrt{2}k_{z} & 0 & 0 & -k_{x}-k_{y} & 0 & k_{x}-k_{y} & 0\\
0 & k_{x}-k_{y} & 0 & k_{x}+k_{y} & 0 & 0 & 0 & 0\\
-k_{x}-k_{y} & 0 & -k_{x}+k_{y} & 0 & 0 & 0 & -\sqrt{2}k_{z} & 0\\
0 & k_{x}+k_{y} & 0 & -k_{x}+k_{y} & \sqrt{2}k_{z} & 0 & 0 & \sqrt{2}k_{z}\\
-k_{x}+k_{y} & 0 & k_{x}+k_{y} & 0 & 0 & -\sqrt{2}k_{z} & 0 & 0
\end{matrix}\right).
\end{align}
For $|v_{s,i}|\ll|v_{p}|$, we neglect the terms proportional to $v_{s,i}$
in Eq. \eqref{eq:HR} and we have $H^{(\text{R})}(\bm{k})\approx\varepsilon_{\text{R}}I_{8}+H_{\bm{k}}^{1\oplus0}\oplus H_{\bm{k}}^{1\oplus0}$
to linear order in $\boldsymbol{k}$. Since the size of the Fermi
sheets around R is large in RhSi, the terms of higher order in $\boldsymbol{k}$
are important to describe the electronic structure near the Fermi
energy. Most importantly, the inclusion of quadratic terms in $\boldsymbol{k}$
results in the band crossing as stated in the main text. Written explicitly,
the $\boldsymbol{k}$-quadratic correction to $H^{(\text{R})}(\bm{k})$
is 
\begin{equation}
\delta H^{(2)}(\boldsymbol{k})\equiv-\frac{v_{2}k^{2}}{2}+v_{1}\delta H_{1,\boldsymbol{k}}^{(\text{R})}+\sum_{i=1,2,3}v_{r,i}\delta H_{r,i,\boldsymbol{k}}^{(\text{R})},
\end{equation}
where 
\begin{align}
\delta H_{1,\boldsymbol{k}}^{(\text{R})} & =\frac{1}{4}\left(\begin{matrix}-k_{x}k_{y} & 0 & 0 & 0 & 0 & \frac{k_{z}(k_{x}-k_{y})}{\sqrt{2}} & 0 & \frac{k_{z}(k_{x}-k_{y})}{\sqrt{2}}\\
0 & 0 & 0 & k_{x}k_{y} & -\frac{k_{z}(k_{x}+k_{y})}{\sqrt{2}} & 0 & \frac{k_{z}(k_{x}-k_{y})}{\sqrt{2}} & 0\\
0 & 0 & k_{x}k_{y} & 0 & 0 & -\frac{k_{z}(k_{x}+k_{y})}{\sqrt{2}} & 0 & \frac{k_{z}(k_{x}+k_{y})}{\sqrt{2}}\\
0 & k_{x}k_{y} & 0 & 0 & -\frac{k_{z}(k_{x}+k_{y})}{\sqrt{2}} & 0 & -\frac{k_{z}(k_{x}-k_{y})}{\sqrt{2}} & 0\\
0 & -\frac{k_{z}(k_{x}+k_{y})}{\sqrt{2}} & 0 & -\frac{k_{z}(k_{x}+k_{y})}{\sqrt{2}} & k_{x}k_{y} & 0 & 0 & 0\\
\frac{k_{z}(k_{x}-k_{y})}{\sqrt{2}} & 0 & -\frac{k_{z}(k_{x}+k_{y})}{\sqrt{2}} & 0 & 0 & 0 & 0 & -k_{x}k_{y}\\
0 & \frac{k_{z}(k_{x}-k_{y})}{\sqrt{2}} & 0 & -\frac{k_{z}(k_{x}-k_{y})}{\sqrt{2}} & 0 & 0 & -k_{x}k_{y} & 0\\
\frac{k_{z}(k_{x}-k_{y})}{\sqrt{2}} & 0 & \frac{k_{z}(k_{x}+k_{y})}{\sqrt{2}} & 0 & 0 & -k_{x}k_{y} & 0 & 0
\end{matrix}\right),\\
\delta H_{r,1,\boldsymbol{k}}^{(\text{R})} & =\frac{1}{4\sqrt{2}}\left(\begin{matrix}0 & k_{z}k_{-} & -ik_{x}k_{y} & -k_{z}k_{+}\\
k_{z}k_{+} & 0 & -k_{z}k_{-} & ik_{x}k_{y}\\
ik_{x}k_{y} & -k_{z}k_{+} & 0 & -k_{z}k_{-}\\
-k_{z}k_{-} & -ik_{x}k_{y} & -k_{z}k_{+} & 0
\end{matrix}\right)\oplus\left(\begin{matrix}0 & k_{z}k_{-} & -ik_{x}k_{y} & -k_{z}k_{+}\\
k_{z}k_{+} & 0 & -k_{z}k_{-} & ik_{x}k_{y}\\
ik_{x}k_{y} & -k_{z}k_{+} & 0 & -k_{z}k_{-}\\
-k_{z}k_{-} & -ik_{x}k_{y} & -k_{z}k_{+} & 0
\end{matrix}\right),\\
\delta H_{r,2,\boldsymbol{k}}^{(\text{R})} & =\frac{1}{4\sqrt{2}}\left(\begin{matrix}0 & -k_{y}k_{z} & 0 & k_{y}k_{z} & \sqrt{2}ik_{x}k_{y} & ik_{x}k_{z} & 0 & -ik_{x}k_{z}\\
-k_{y}k_{z} & 0 & -k_{y}k_{z} & 0 & -ik_{x}k_{z} & 0 & ik_{x}k_{z} & \sqrt{2}ik_{x}k_{y}\\
0 & -k_{y}k_{z} & 0 & -k_{y}k_{z} & 0 & -ik_{x}k_{z} & -\sqrt{2}ik_{x}k)y & -ik_{x}k_{z}\\
k_{y}k_{z} & 0 & -k_{y}k_{z} & 0 & ik_{x}k_{z} & \sqrt{2}ik_{x}k_{y} & ik_{x}k_{z} & 0\\
-\sqrt{2}ik_{x}k_{y} & ik_{x}k_{z} & 0 & -ik_{x}k_{z} & 0 & k_{y}k_{z} & 0 & -k_{y}k_{z}\\
-ik_{x}k_{z} & 0 & ik_{x}k_{z} & \sqrt{2}ik_{x}k_{y} & k_{y}k_{z} & 0 & k_{y}k_{z} & 0\\
0 & -ik_{x}k_{z} & -\sqrt{2}ik_{x}k)y & -ik_{x}k_{z} & 0 & -k_{y}k_{z} & 0 & k_{y}k_{z}\\
ik_{x}k_{z} & \sqrt{2}ik_{x}k_{y} & ik_{x}k_{z} & 0 & -k_{y}k_{z} & 0 & k_{y}k_{z} & 0
\end{matrix}\right),\\
\delta H_{r,3,\bm{k}}^{(\text{R})} & =\frac{1}{8}\left(\begin{matrix}-k_{x}^{2}-k_{z}^{2} & 0 & -k_{x}^{2}+k_{z}^{2} & 0\\
0 & -2k_{y}^{2} & 0 & 0\\
-k_{x}^{2}+k_{z}^{2} & 0 & -k_{x}^{2}-k_{z}^{2} & 0\\
0 & 0 & 0 & 2(k_{x}^{2}+k_{y}^{2}+k_{z}^{2})
\end{matrix}\right)\oplus\left(\begin{matrix}-k_{x}^{2}-k_{z}^{2} & 0 & -k_{x}^{2}+k_{z}^{2} & 0\\
0 & -2k_{y}^{2} & 0 & 0\\
-k_{x}^{2}+k_{z}^{2} & 0 & -k_{x}^{2}-k_{z}^{2} & 0\\
0 & 0 & 0 & 2(k_{x}^{2}+k_{y}^{2}+k_{z}^{2})
\end{matrix}\right).
\end{align}
Note that only $\delta H_{1,\boldsymbol{k}}^{(\text{R})}$ and $\delta H_{r,2,\boldsymbol{k}}^{(\text{R})}$
lift the artificial degeneracy in the model $H_{\boldsymbol{k}}^{1\oplus0}\oplus H_{\boldsymbol{k}}^{1\oplus0}$.
Given $v_{1}\gg|v_{r,i}|$, it is sufficient to keep only $v_{1}$
and $v_{2}$ while neglecting $v_{r,i}$'s to capture not only the
qualitative but also quantitative aspects of the band structure around
$\text{R}$. In Fig. \ref{fig:figS1}, the band structures from several
continuum models are shown in comparison to the band structure from
the tight-binding model. Comparing Figs. \ref{fig:figS1}(a) and \ref{fig:figS1}(b),
no significant modification from neglecting $v_{r,i}$ occurs, and
especially the location of the band crossing point is well predicted
even without $v_{r,i}$'s. Note that the $\boldsymbol{k}$-linear $H_{\boldsymbol{k}}^{1\oplus0}\oplus H_{\boldsymbol{k}}^{1\oplus0}$
alone does not capture the band crossing near R, as shown in Fig.~\ref{fig:figS1}(c).
Hence, we conclude that $-v_{2}k^{2}/2-v_{1}\delta H_{1,\boldsymbol{k}}^{(\text{R})}$
is the major $\boldsymbol{k}$-quadratic correction to $H^{(\text{R})}(\bm{k})\approx\varepsilon_{\text{R}}I_{8}+H_{\bm{k}}^{1\oplus0}\oplus H_{\bm{k}}^{1\oplus0}$.

\section{Gap functions}

In the main text, we consider only the short-range interactions which
result in the short-range Cooper pairings: the pairings between electrons
at the same site or the pairings between electrons at the neighboring
sites. Focusing on the time-reversal invariant gap functions respecting
all the spatial symmetries of RhSi, the gap function $\Delta(\boldsymbol{k})$
can be represented by a linear combination of $M_{i,\boldsymbol{k}}$'s
with $i=0,1,2,\cdots,8$: 
\begin{equation}
\Delta(\boldsymbol{k})=\sum_{i=0}^{8}\Delta_{i}M_{i,\boldsymbol{k}}.\label{eq:Delta_k_Lattice}
\end{equation}
The basis of the representation is same with that used in Eq.~(7)
in the main text. Among the 9 matrices, $M_{0,\boldsymbol{k}}$, $M_{1,\boldsymbol{k}}$,
and $M_{5,\boldsymbol{k}}$ are the matrices independent of the real
spin, and thus represent the real-spin-singlet pairings, while the
other 6 matrices denote the real-spin-triplet pairings. Considering
the symmetric/antisymmetric character of orbitals in addition to even/odd
parity in $\boldsymbol{k}$ and singlet/triplet in spin degrees of
freedom, symmetry transformation properties of $M_{i,\boldsymbol{k}}$
are summarized in Table. \ref{tab:Gap-symmetry}. Note that $M_{2,\boldsymbol{k}}$,
$M_{3,\boldsymbol{k}}$, and $M_{4,\boldsymbol{k}}$ are all even-parity,
spin-triplet, and orbital-antisymmetric gap functions which correspond
to the $s_{+}\oplus s_{-}$ gap functions as shown in the following
subsections.

\begin{table}[H]
\begin{centering}
\begin{tabular}{|c|c|c|c|c|}
\hline 
\multicolumn{1}{|c|}{} & $\boldsymbol{k}$ & Spin & Orbital & Examples\tabularnewline
\hline 
\hline 
Type 1 & even & singlet & symmetric & $M_{0},\;M_{1}$\tabularnewline
\hline 
Type 2 & even & triplet & antisymmetric & $M_{2},\;M_{3},\;M_{4}$\tabularnewline
\hline 
Type 3 & odd & singlet & symmetric & $M_{5}$\tabularnewline
\hline 
Type 4 & odd & triplet & antisymmetric & $M_{6},\;M_{7},\;M_{8}$\tabularnewline
\hline 
\end{tabular}
\par\end{centering}
\caption{\label{tab:Gap-symmetry}Symmetry transformation properties of momentum
($\boldsymbol{k}$), orbital ($\tau_{i}\mu_{j}$), and spin ($s_{i}$)
degrees of freedom under the fermion exchange. For the expression
of $M_{i}$ by Pauli matrices $\tau_{i}$, $\mu_{i}$ and $s_{i}$,
see Eq. (3).}
\end{table}

\subsection{Expansion of $M_{i,\bm{k}}$ near the $\Gamma$ point}

Writing down the first non-zero term, the Taylor expansion of $M_{i,\bm{k}}^{(\Gamma)}\equiv{\cal \mathbb{P}}_{\Gamma}^{\dagger}M_{i,\bm{k}}{\cal \mathbb{P}}_{\Gamma}$
around $\Gamma$ in $\boldsymbol{k}$ is given by
\begin{equation}
\begin{array}{cccc}
M_{0,\bm{k}}^{(\Gamma)}=+I_{6},\hfill & M_{1,\bm{k}}^{(\Gamma)}=-I_{6},\hfill & M_{2,\bm{k}}^{(\Gamma)}=\Delta_{\pm}^{\Gamma},\hfill & M_{3,\bm{k}}^{(\Gamma)}=-\Delta_{\pm}^{\Gamma},\\
M_{6,\bm{k}}^{(\Gamma)}=\delta H_{s,1}^{(\Gamma)}(\boldsymbol{k}), & M_{7,\bm{k}}^{(\Gamma)}=\delta H_{s,2}^{(\Gamma)}(\boldsymbol{k}), & M_{8,\bm{k}}^{(\Gamma)}=\delta H_{s,3}^{(\Gamma)}(\boldsymbol{k}),
\end{array}
\end{equation}
and
\begin{align}
M_{4,\bm{k}}^{(\Gamma)}= & \left(\begin{array}{cccccc}
0 & \frac{(k_{x}-ik_{y})k_{z}}{4\sqrt{3}} & -\frac{ik_{x}k_{y}}{4\sqrt{3}} & 0 & -\frac{(k_{x}+2ik_{y})k_{z}}{4\sqrt{6}} & -\frac{ik_{x}k_{y}}{4\sqrt{6}}\\
\frac{(k_{x}+ik_{y})k_{z}}{4\sqrt{3}} & 0 & 0 & -\frac{ik_{x}k_{y}}{4\sqrt{3}} & \frac{ik_{x}k_{y}}{4\sqrt{2}} & -\frac{k_{x}k_{z}}{4\sqrt{2}}\\
\frac{ik_{x}k_{y}}{4\sqrt{3}} & 0 & 0 & -\frac{(k_{x}-ik_{y})k_{z}}{4\sqrt{3}} & -\frac{k_{x}k_{z}}{4\sqrt{2}} & \frac{ik_{x}k_{y}}{4\sqrt{2}}\\
0 & \frac{ik_{x}k_{y}}{4\sqrt{3}} & -\frac{(k_{x}+ik_{y})k_{z}}{4\sqrt{3}} & 0 & -\frac{ik_{x}k_{y}}{4\sqrt{6}} & -\frac{(k_{x}-2ik_{y})k_{z}}{4\sqrt{6}}\\
-\frac{(k_{x}-2ik_{y})k_{z}}{4\sqrt{6}} & -\frac{ik_{x}k_{y}}{4\sqrt{2}} & -\frac{k_{x}k_{z}}{4\sqrt{2}} & \frac{ik_{x}k_{y}}{4\sqrt{6}} & 0 & 0\\
\frac{ik_{x}k_{y}}{4\sqrt{6}} & -\frac{k_{x}k_{z}}{4\sqrt{2}} & -\frac{ik_{x}k_{y}}{4\sqrt{2}} & -\frac{(k_{x}+2ik_{y})k_{z}}{4\sqrt{6}} & 0 & 0
\end{array}\right),\\
M_{5,\bm{k}}^{(\Gamma)}= & \left(\begin{array}{cccccc}
\frac{k_{z}}{2} & \frac{k_{x}-ik_{y}}{2\sqrt{3}} & 0 & 0 & -\frac{k_{x}-ik_{y}}{2\sqrt{6}} & 0\\
\frac{k_{x}+ik_{y}}{2\sqrt{3}} & \frac{k_{z}}{6} & \frac{k_{x}-ik_{y}}{3} & 0 & \frac{k_{z}}{3\sqrt{2}} & -\frac{k_{x}-ik_{y}}{6\sqrt{2}}\\
0 & \frac{k_{x}+ik_{y}}{3} & -\frac{k_{z}}{6} & \frac{k_{x}-ik_{y}}{2\sqrt{3}} & \frac{k_{x}+ik_{y}}{6\sqrt{2}} & \frac{k_{z}}{3\sqrt{2}}\\
0 & 0 & \frac{k_{x}+ik_{y}}{2\sqrt{3}} & -\frac{k_{z}}{2} & 0 & \frac{k_{x}+ik_{y}}{2\sqrt{6}}\\
-\frac{k_{x}+ik_{y}}{2\sqrt{6}} & \frac{k_{z}}{3\sqrt{2}} & \frac{k_{x}-ik_{y}}{6\sqrt{2}} & 0 & \frac{k_{z}}{3} & \frac{k_{x}-ik_{y}}{3}\\
0 & -\frac{k_{x}+ik_{y}}{6\sqrt{2}} & \frac{k_{z}}{3\sqrt{2}} & \frac{k_{x}-ik_{y}}{2\sqrt{6}} & \frac{k_{x}+ik_{y}}{3} & -\frac{k_{z}}{3}
\end{array}\right).
\end{align}
Note that $M_{0-3,\bm{k}}^{(\Gamma)}$ are non-zero at $\bm{k}=0$,
whereas the other gap functions are zero at $\bm{k}=0$. Considering
that the Fermi sheets around $\Gamma$ is small in RhSi, only the
gap functions $M_{0-3,\bm{k}}^{(\Gamma)}$ are able to open a sizable
superconducting gap over the Fermi sheets near $\Gamma$. Hence, around
$\Gamma$, the gap function $\Delta(\boldsymbol{k})$ can be approximated
as 
\begin{equation}
\Delta(\boldsymbol{k})\approx(\Delta_{0}-\Delta_{1})I_{6}+(\Delta_{2}-\Delta_{3})\Delta_{\pm}^{\Gamma}.
\end{equation}

\subsection{Expansion of $M_{i,\bm{k}}$ near the ${\rm R}$ point}

Writing down the first non-zero term, the Taylor expansion of $M_{i,\bm{k}}^{(\text{R})}\equiv{\cal \mathbb{P}}_{\text{R}}^{\dagger}M_{i,\boldsymbol{k}_{\text{R}}+\bm{k}}{\cal \mathbb{P}}_{\text{R}}$
around ${\rm R}$ in $\boldsymbol{k}$ is given by
\begin{equation}
\begin{array}{ccc}
M_{0,\boldsymbol{k}}^{(\text{R})}=I_{8},\hfill & M_{4,\boldsymbol{k}}^{(\text{R})}=\Delta_{\pm}^{\text{R}},\hfill\\
M_{1,\bm{k}}^{(\text{R})}=\delta H_{1,\boldsymbol{k}}^{(\text{R})},\hfill & M_{2,\bm{k}}^{(\text{R})}=\delta H_{r,1,\boldsymbol{k}}^{(\text{R})},\hfill & M_{3,\bm{k}}^{(\text{R})}=\delta H_{r,2,\boldsymbol{k}}^{(\text{R})},\hfill\\
M_{6,\bm{k}}^{(\text{R})}=\delta H_{s,1,\boldsymbol{k}}^{(\text{R})},\hfill & M_{7,\bm{k}}^{(\text{R})}=\delta H_{s,2,\boldsymbol{k}}^{(\text{R})},\hfill & M_{8,\bm{k}}^{(\text{R})}=\delta H_{s,3,\boldsymbol{k}}^{(\text{R})},\hfill
\end{array}
\end{equation}
and
\begin{equation}
M_{5,\bm{k}}^{(\text{R})}=-\frac{1}{2\sqrt{2}}\left(\begin{matrix}\sqrt{2}k_{z} & k_{x}-ik_{y} & 0 & k_{x}-ik_{y}\\
k_{x}+ik_{y} & 0 & k_{x}-ik_{y} & -\sqrt{2}k_{z}\\
0 & k_{x}+ik_{y} & -\sqrt{2}k_{z} & -k_{x}-ik_{y}\\
k_{x}+ik_{y} & -\sqrt{2}k_{z} & -k_{x}+ik_{y} & 0
\end{matrix}\right)\oplus\left(\begin{matrix}\sqrt{2}k_{z} & k_{x}-ik_{y} & 0 & k_{x}-ik_{y}\\
k_{x}+ik_{y} & 0 & k_{x}-ik_{y} & -\sqrt{2}k_{z}\\
0 & k_{x}+ik_{y} & -\sqrt{2}k_{z} & -k_{x}-ik_{y}\\
k_{x}+ik_{y} & -\sqrt{2}k_{z} & -k_{x}+ik_{y} & 0
\end{matrix}\right).
\end{equation}
Note that only $M_{0,\boldsymbol{k}}^{(\text{R})}$ and $M_{4,\bm{k}}^{(\text{R})}$
are non-zero at $\bm{k}=0$, and these gap functions are the decisive
ones determining the sign of the gap function for each band for small
$k$. Therefore, near R, we can approximate $\Delta(\boldsymbol{k})$
in Eq. \eqref{eq:Delta_k_Lattice} as 
\begin{equation}
\Delta(\boldsymbol{k})\approx\Delta_{0}I_{8}+\Delta_{4}\Delta_{\pm}^{\text{R}}
\end{equation}
in the small $k$ limit. However, when the Fermi sheets are large,
the typical magnitude of the Fermi momentum is too large to ignore
the $\boldsymbol{k}$-dependent contributions from other $M_{i,\boldsymbol{k}}^{(\text{R})}$'s.
The superconducting phases with non-negligible $\Delta_{i\neq0,4}$'s
will be discussed in Sec. \ref{sec:V}.

\section{Generic DIII Topological Superconducting phases with the $s_{+}\oplus s_{-}$
gap functions\label{sec:Topological phase}}

In the main text, we focus on the topological superconducting phase
with a MZM on $\bar{\Gamma}$ and nodal rings around R because $\mu_{\text{R}}>E_{\text{c,2}}$
in the candidate materials such as RhSi. If the nodal rings are not
present, the DIII TSC phases from $\Delta_{\text{\ensuremath{\Gamma(\text{R})}}}(\beta_{\Gamma(\text{R})})$
with a higher winding number $|w_{\text{3D}}|>1$ are possible in
which multiple topologically stable MZMs are guaranteed.

Table \ref{tab:tabS1} shows all possible $w_{\text{3D}}$ in the
fully gapped superconducting phases from $\Delta_{\Gamma}(\beta_{\Gamma})$
and $\Delta_{\text{R}}(\beta_{\text{R}})$ for non-zero $\alpha_{\text{R}}$
assuming that $\alpha_{\Gamma}$ is positive. For a negative $\alpha_{\Gamma}$,
we can obtain the corresponding winding numbers from $(|\alpha_{\Gamma}|,-\alpha_{\text{R}})$.
From the table, one can find that TSC phases with a higher winding
number $w_{\text{3D}}$ can be realized from $\Delta_{\text{\ensuremath{\Gamma(\text{R})}}}(\beta_{\Gamma(\text{R})})$.

\begin{table}[H]
\begin{centering}
\begin{tabular}{cc|cc}
\hline 
(a) &  & \multicolumn{2}{c}{$(w_{\text{3D,R}},w_{\text{1D,R}})$}\tabularnewline
 &  & $(-2,+1)$ & $(2,+1)$\tabularnewline
\hline 
\hline 
\multirow{3}{*}[-25pt]{\begin{turn}{90}
\begin{minipage}[b]{10pt}%
\makebox[35pt]{%
$(w_{\text{3D,}\Gamma},w_{\text{1D,}\Gamma})$%
}%
\end{minipage}
\end{turn}} & %
\begin{minipage}[c][21pt]{35pt}%
\makebox[35pt]{%
$(2,+1)$%
}%
\end{minipage} & $(0,+1)$ & $(4,+1)$\tabularnewline
\cline{2-4} \cline{3-4} \cline{4-4} 
 & %
\begin{minipage}[c][21pt]{35pt}%
\makebox[35pt]{%
$(1,-1)$%
}%
\end{minipage} & $(-1,-1)$ & $(3,-1)$\tabularnewline
\cline{2-4} \cline{3-4} \cline{4-4} 
 & %
\begin{minipage}[c][21pt]{35pt}%
\makebox[35pt]{%
$(-2,+1)$%
}%
\end{minipage} & $(-4,+1)$ & $(0,+1)$\tabularnewline
\hline 
\end{tabular}\quad{}\quad{}\quad{}\quad{}%
\begin{tabular}{cc|cc}
\hline 
(b) &  & \multicolumn{2}{c}{$(w_{\text{3D,R}},w_{\text{1D,R}})$}\tabularnewline
 &  & $(2,+1)$ & $(-2,+1)$\tabularnewline
\hline 
\hline 
\multirow{3}{*}[-25pt]{\begin{turn}{90}
\begin{minipage}[b]{10pt}%
\makebox[35pt]{%
$(w_{\text{3D,}\Gamma},w_{\text{1D,}\Gamma})$%
}%
\end{minipage}
\end{turn}} & %
\begin{minipage}[c][21pt]{35pt}%
\makebox[35pt]{%
$(2,+1)$%
}%
\end{minipage} & $(4,+1)$ & $(0,+1)$\tabularnewline
\cline{2-4} \cline{3-4} \cline{4-4} 
 & %
\begin{minipage}[c][21pt]{35pt}%
\makebox[35pt]{%
$(1,-1)$%
}%
\end{minipage} & $(3,-1)$ & $(-1,-1)$\tabularnewline
\cline{2-4} \cline{3-4} \cline{4-4} 
 & %
\begin{minipage}[c][21pt]{35pt}%
\makebox[35pt]{%
$(-2,+1)$%
}%
\end{minipage} & $(0,+1)$ & $(-4,+1)$\tabularnewline
\hline 
\end{tabular}
\par\end{centering}
\caption{\label{tab:tabS1}Winding numbers $(w_{\text{3D}},w_{\text{1D}})$
for (a) $\alpha_{\text{R}}>0$ and (b) $\alpha_{\text{R}}<0$ with
$\alpha_{\Gamma}>0$. Here we arrange possible $w_{\text{3D,}\Gamma}$
($w_{\text{3D,R}}$) from the bottom to top (left to right) with increasing
$\beta_{\Gamma}$ ($\beta_{\text{R}}$).}
\end{table}

\section{Nodal rings around $\text{R}$ and their stability \label{sec:Nodal-rings-aroundR}}

As studied in the main text, topological nodal rings around $\text{R}$
in the BdG spectrum arise when $\Delta_{\pm}^{\text{R}}$ is dominant
in $\Delta_{\text{R}}(\beta_{\text{R}})$ and the Fermi level $\mu_{\text{R}}$
lies above the band crossing energy level $E_{c,2}$. In this section,
we delineate the nodal rings in the BdG spectrum and their topological
stability.

\begin{figure}[t]
\includegraphics[width=0.95\textwidth]{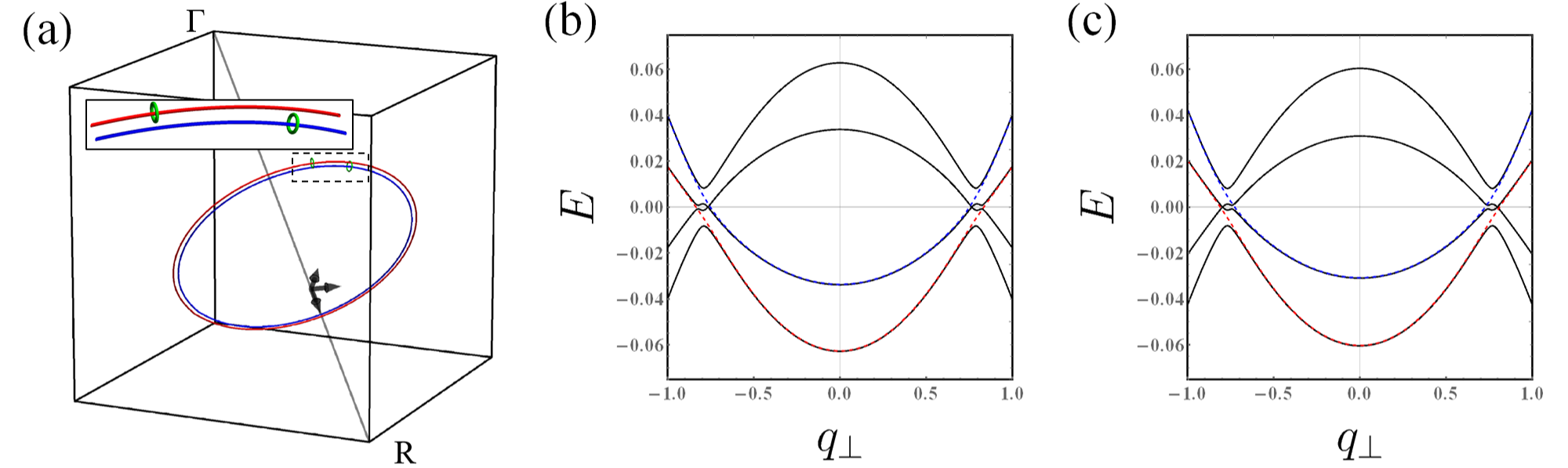}

\caption{\label{fig:figS2}\hspace{0pt}(a) $(q_{1},q_{2},q_{3}$) coordinate
system, nodal rings and the integration paths ${\cal L}_{n}$ in the
first Brillouin zone. The blue (red) ring represents the nodal ring
obtained from the solution $\boldsymbol{q}^{(1)}$ ($\boldsymbol{q}^{(2)}$).
The green rings represent ${\cal L}_{n}$. (b), (c) BdG spectra at
$q_{3}=q_{3}^{(1)}$ and $q_{3}=q_{3}^{(2)}$. The black lines represent
the BdG spectra while the blue/red dashed lines show the two normal
electronic bands crossing at $E=E_{c,2}$.}
\end{figure}

To construct an effective Hamiltonian for the bands crossing at $E=E_{c,2}$,
we first note that, on the $\Gamma-\text{R}$ line which is invariant
under $C_{3,[111]}$, the three-fold rotaion along the {[}111{]} direction,
the two bands which cross at $E=E_{c,2}$ have different eigenvalues
$\lambda_{1}=-1$ and $\lambda_{2}=\exp(-i\pi/3)$ under $C_{3,[111]}$
which protects the band crossing point. Using the wave functions for
these bands as the basis vectors, the effective Hamiltonian around
the crossing point is written as \citep{Fang2012} 
\begin{align}
H_{\text{BdG}}(\boldsymbol{q}) & =\left(\begin{matrix}H(\boldsymbol{q}) & \boldsymbol{\Delta}\\
\boldsymbol{\Delta} & -H(\boldsymbol{q})
\end{matrix}\right),\label{eq:NodalRingHamiltonian}
\end{align}
with 
\begin{align}
H(\boldsymbol{q}) & =h_{0}(\boldsymbol{q})\sigma_{0}+\boldsymbol{h}(\boldsymbol{q})\cdot\boldsymbol{\sigma},\\
h_{0}(\boldsymbol{q}) & =E_{c,2}-\mu+c_{\parallel}q_{3}+c_{\perp}q_{\perp}^{2},\\
h_{3}(\boldsymbol{q}) & =a_{\parallel}q_{3}+b_{\perp}q_{\perp}^{2},\\
h_{1(2)}(\boldsymbol{q}) & =a_{\perp}q_{1(2)},\\
\boldsymbol{\Delta} & =\left(\begin{matrix}\Delta_{1} & 0\\
0 & -\Delta_{2}
\end{matrix}\right)+O(q^{2}),
\end{align}
where $q_{\perp}^{2}=q_{1}^{2}+q_{2}^{2}$. As shown in Fig. \ref{fig:figS2}(a),
we define the band crossing point as the origin of $\boldsymbol{q}=(q_{1},q_{2},q_{3})$.
Here, the $q_{3}$ axis is chosen along the $\Gamma-\text{R}$ direction,
and $q_{1}$, $q_{2}$ are the in-plane coordinates perpendicular
to the $q_{3}$ axis. Note that $\Delta_{1}\Delta_{2}>0$ when $\Delta_{\pm}^{(\text{R})}$
is dominant in which the nodal rings can appear. Hence, from now on
we assume that both $\Delta_{1}$ and $\Delta_{2}$ are positive without
loss of generality.

To get an insight from the weak-pairing limit, we rewrite $H_{\text{BdG}}(\boldsymbol{q})$
in the basis which diagonalizes $H(\boldsymbol{q})$: 
\begin{equation}
H_{\text{BdG}}(\boldsymbol{q})\approx\left(\begin{matrix}h_{0}+h & 0 & \delta+\frac{h_{3}}{h}\bar{\Delta} & \frac{h_{1}-ih_{2}}{h}\bar{\Delta}\\
0 & h_{0}-h & \frac{h_{1}+ih_{2}}{h}\bar{\Delta} & \delta-\frac{h_{3}}{h}\bar{\Delta}\\
\delta+\frac{h_{3}}{h}\bar{\Delta} & \frac{h_{1}-ih_{2}}{h}\bar{\Delta} & -h_{0}-h & 0\\
\frac{h_{1}+ih_{2}}{h}\bar{\Delta} & \delta-\frac{h_{3}}{h}\bar{\Delta} & 0 & -h_{0}+h
\end{matrix}\right),
\end{equation}
where $h=|\boldsymbol{h}|=\sqrt{h_{1}^{2}+h_{2}^{2}+h_{3}^{2}}$,
$\bar{\Delta}=(\Delta_{1}+\Delta_{2})/2$, and $\delta=(\Delta_{1}-\Delta_{2})/2$.
In the weak pairing limit where $\Delta_{1}$ and $\Delta_{2}$ are
very small, the gap function for the electronic band with the dispersion
$h_{0}+h$ is $\delta+\frac{h_{3}}{h}\bar{\Delta}$, which ranges
from $-\Delta_{2}$ to $+\Delta_{1}$. Therefore, we could expect
a region in which the BdG spectrum is gapless. Exactly the same argument
is applicable to the BdG spectrum from the electronic band with the
dispersion $h_{0}-h$.

To confirm the existence of the gapless excitation in the BdG spectrum,
we diagonalize $H_{\text{BdG}}(\boldsymbol{q})$. The eigenvalues
of $H_{\text{BdG}}(\boldsymbol{q})$ are given by 
\begin{align}
E_{\pm}^{2}= & h_{0}^{2}+h^{2}+\bar{\Delta}^{2}+\delta^{2}\pm2\sqrt{(h_{1}^{2}+h_{2}^{2})^{2}(h_{0}^{2}+\bar{\Delta}^{2})+(h_{0}h_{3}+\bar{\Delta}\delta)^{2}}.
\end{align}
$E_{-}^{2}(\boldsymbol{q})=0$ occurs at $\boldsymbol{q}$ satisfying
\begin{equation}
h_{0}^{2}(\boldsymbol{q})-h^{2}(\boldsymbol{q})=\delta^{2}-\bar{\Delta}^{2},\quad h_{3}(\boldsymbol{q})\bar{\Delta}=h_{0}(\boldsymbol{q})\delta.\label{eq:NodalRing}
\end{equation}
In the weak-pairing limit, the solutions $\boldsymbol{q}$ for the
first equation form the two Fermi sheets around the band crossing
point, whereas those for the second equation form paraboloids around
the $q_{3}$-axis in general. Therefore, the solutions of Eq. \eqref{eq:NodalRing}
are found near the intersections between the paraboloids and the Fermi
sheets.

Considering the isotropy in $(q_{1},q_{2})$ in $H_{\text{BdG}}(\boldsymbol{q})$,
we find two solutions $\boldsymbol{q}^{(1)}=(q_{\perp}^{(1)},q_{3}^{(1)})$
and $\boldsymbol{q}^{(2)}=(q_{\perp}^{(2)},q_{3}^{(2)})$ of Eq. \eqref{eq:NodalRing}:
\begin{align}
(q_{\perp}^{(1)},q_{3}^{(1)})\approx(0.766,-0.337) & ,\quad(q_{\perp}^{(2)},q_{3}^{(2)})\approx(0.796,-0.343).
\end{align}
Here, the parameters $c_{\perp}=0.077$, $c_{\parallel}=-0.45$, $a_{\perp}=0.006$,
$a_{\parallel}=-0.043$, $b_{\perp}=-0.024$ eV are used for RhSi,
and $\Delta_{1}=0.004$, $\Delta_{2}=0.003$ and $E_{c,2}-\mu=-0.2$
eV are chosen as an example.

Figures \ref{fig:figS2}(b) and \ref{fig:figS2}(c) show the BdG spectrum
of $H_{\text{BdG}}(\boldsymbol{q})$ along the $q_{\perp}$-axis at
$q_{3}=q_{3}^{(1)}$ and $q_{3}=q_{3}^{(2)}$, respectively. The nodal
points are clearly seen. These points actually correspond to nodal
rings in Fig. \ref{fig:figS2}(a).

Knowing the location of nodal rings, it is straightforward to test
the stability of the nodal rings. Calculating the winding number \citep{Schnyder2011}
$w_{\text{1D}}^{\text{(NR)}}=\oint_{{\cal L}_{n}}\frac{\text{d}\phi}{2\pi i}\text{Tr}\left[Q_{n}^{-1}\partial_{\theta}Q_{n}\right]$
along the paths ${\cal L}_{n}$ drawn as the green rings in the inset
of Fig. \ref{fig:figS2}(a) where $Q_{n}(r,\theta)=H(q_{\perp}^{(n)}+r\cos\theta,0,q_{3}^{(n)}+r\sin\theta)-i\boldsymbol{\Delta}$,
we get $w_{\text{1D}}^{\text{(NR)}}=+1$ for $n=1$ and $w_{\text{1D}}^{\text{(NR)}}=-1$
for $n=2$. Thus, the nodal rings in the BdG spectrum are topologically
stable.

Two remarks are in order regarding the description of the nodal rings
presented here. First, even if the Hamiltonian $H_{\text{BdG}}(\boldsymbol{q})$
in Eq. \eqref{eq:NodalRingHamiltonian} is valid around the band crossing
point, the nodal rings could survive beyond the range of validity
of $H_{\text{BdG}}(\boldsymbol{q})$ due to their topological stability.
Second, the argument given here is also applicable to the system with
an anti-crossing between the bands. The only modification is just
to add a mass term to $h_{1}(\boldsymbol{q})$ or $h_{2}(\boldsymbol{q})$
to describe the anti-crossing. This addition may affect the location
of nodal rings but does not invalidate their existence.

\section{Linearized gap equation\label{sec:V}}

\subsection{Calculation of the linearized gap equation }

To estimate the superconducting gap functions from the three short-ranged
interactions introduced in the main text, we need to find the minimum
point of the free energy ${\cal F}[\hat{\Delta}_{i}]$ \citep{Bauer2012,Suh2019}
where $\hat{\Delta}_{i}$'s denote the pairing channels. Written up
to the second order in $\hat{\Delta}_{i}$, the free energy is given
by 
\begin{equation}
\mathcal{F}[\hat{\Delta}_{i}]=\frac{1}{2}\sum_{i}\frac{1}{U_{i}}\text{Tr}\left[\hat{\Delta}_{i}^{\dagger}\hat{\Delta}_{i}\right]-\frac{k_{B}T}{2}\sum_{\boldsymbol{k},\omega_{n},i,j}\text{Tr}\left[\hat{\Delta}_{i}^{\dagger}G_{0}(i\omega_{n},\boldsymbol{k})\hat{\Delta}_{j}G_{0}(-i\omega_{n},\boldsymbol{k})\right],\label{eq:Free energy}
\end{equation}
where $U_{i}$ is the interaction strength for the $i$-th pairing
channel $\hat{\Delta}_{i}$, $G_{0}$ is the Green's function for
a quasi-particle in the normal phase and $\omega_{n}$ is a fermionic
Matsubara frequency. To our case, the pairing channel $\hat{\Delta}_{i}$
is identified with $\Delta_{i}M_{i,\boldsymbol{k}}$ and the interaction strengths
are defined as $U_{0}=U/4$, $U_{1,5}=V-3J$, and $U_{2,3,4,6,7,8}=V+J$,
as stated in the main text. To find the minimum point of ${\cal F}[\hat{\Delta}_{i}]$,
we first differentiate it with respect to $\Delta_{i}$ yielding a
set of equations named the linearized gap equation (LGE). Written
in the matrix form, the LGE is given by 
\begin{align}
\Delta & =U\chi\Delta,\label{eq:LGE}
\end{align}
where $U=\text{diag}(U_{0},U_{1},\dots,U_{8})$, $\Delta=(\Delta_{0},\Delta_{1},\dots,\Delta_{8})$,
and $\chi$ is the matrix of the pairing susceptibility whose elements
are defined as 
\begin{equation}
\chi_{ij}=-k_{B}T\text{ }\sum_{\omega_{n},\boldsymbol{k}}\text{Tr}\left[M_{i,\boldsymbol{k}}G_{0}(i\omega_{n},\boldsymbol{k})M_{j,\boldsymbol{k}}G_{0}(-i\omega_{n},\boldsymbol{k})\right].
\end{equation}
Completing the summation over Matsubara frequencies, we have 
\begin{align}
\chi_{ij}= & -\sum_{\xi,\xi'}\sum_{\boldsymbol{k}}\frac{\langle\xi,\boldsymbol{k}|M_{i,\boldsymbol{k}}|\xi',\boldsymbol{k}\rangle\langle\xi',\boldsymbol{k}|M_{j,\boldsymbol{k}}|\xi,\boldsymbol{k}\rangle}{\varepsilon_{\xi,\boldsymbol{k}}+\varepsilon_{\xi',\boldsymbol{k}}}\left[\tanh\frac{\beta\varepsilon_{\xi,\boldsymbol{k}}}{2}+\tanh\frac{\beta\varepsilon_{\xi',\boldsymbol{k}}}{2}\right],
\end{align}
where $\xi,\xi'$ are the band indices. Changing the summation over
$\boldsymbol{k}$ to the integration over energy for each band, only
the bands in the vicinity of the Fermi sheets contribute to $\chi_{ij}$.
Thus we obtain 
\begin{align}
\chi_{ij} & =-\sum_{\xi}\sum_{\varepsilon_{\xi'}=\varepsilon_{\xi}}\left[\int_{\text{FS}_{\xi}}\frac{\text{d}^{2}\boldsymbol{k}}{v_{\text{F}}}\langle\xi,\boldsymbol{k}|M_{i,\boldsymbol{k}}|\xi',\boldsymbol{k}\rangle\langle\xi',\boldsymbol{k}|M_{j,\boldsymbol{k}}|\xi,\boldsymbol{k}\rangle\right]\log\frac{2e^{\gamma}\beta\varepsilon_{c}}{\pi},
\end{align}
where we assume that the system is time-reversal symmetric or inversion
symmetric. Here $\gamma\sim0.577$ is the Euler's constant, $v_{\text{F}}\equiv\lVert\nabla\varepsilon_{\text{F}}\rVert$
is the Fermi velocity and $\varepsilon_{c}$ is the energy cutoff
for the pairing interactions, which is assumed to be the same for
all bands and all interactions. By finding the solution $\Delta$
of Eq. \eqref{eq:LGE} with the largest positive eigenvalue of $U\bar{\chi}$
where 
\begin{equation}
\bar{\chi}_{ij}\equiv-\sum_{\xi}\sum_{\varepsilon_{\xi'}=\varepsilon_{\xi}}\left[\int_{\text{FS}_{\xi}}\frac{\text{d}^{2}\boldsymbol{k}}{v_{\text{F}}}\langle\xi,\boldsymbol{k}|M_{i,\boldsymbol{k}}|\xi',\boldsymbol{k}\rangle\langle\xi',\boldsymbol{k}|M_{j,\boldsymbol{k}}|\xi,\boldsymbol{k}\rangle\right],
\end{equation}
we can estimate the most favorable superconducting gap function with
the highest transition temperature.

For the practical calculation of $\bar{\chi}_{ij}$, we first decompose
$\bar{\chi}_{ij}$ into the contributions from $\Gamma$ and R as
$\bar{\chi}_{ij}=\bar{\chi}_{ij}^{(\Gamma)}+\bar{\chi}_{ij}^{(\text{R})}$
and make two approximations. First, we replace the integration over
the Fermi sheet $\int_{\text{FS}_{\xi}}\text{d}^{2}\boldsymbol{k}$
by $\int\text{d}\theta\text{d}\phi k_{\text{F}}^{2}(\theta,\phi)\sin\theta$
since the shape of the Fermi sheets around $\Gamma$ and R looks very
close to that of a sphere. Small distortion from the sphere is taken
into account by allowing the Fermi momentum $k_{\text{F}}$ to be
a function of $\theta$ and $\phi$. Second, we ignore the interband
pairings, which is acceptable in the weak pairing limit. One could
concern with the fact that there are symmetry-protected double degeneracy
on the $k_{x,y,z}=\pi$ planes as the energy difference between degenerate
bands is smaller than the magnitude of superconducting gap. However,
the intersections between the Fermi sheets and those planes are just
one-dimensional loops, whose contribution from the interband pairings
to the surface integral over the Fermi sheets is usually negligible.
These approximations lead to the following simplified formulas: 
\begin{align}
\bar{\chi}_{ij}^{(\Gamma)}= & -\sum_{\xi\in\left\{ \text{in},\text{out}\right\} }\left[\int_{\text{FS}_{\xi}}\text{d}\theta\text{d}\phi\frac{k_{\text{F},\Gamma}^{2}(\theta,\phi)\text{sin}\theta}{v_{\text{F},\Gamma}(\theta,\phi)}\langle\xi,\boldsymbol{k}|M_{i,\boldsymbol{k}}|\xi,\boldsymbol{k}\rangle\langle\xi,\boldsymbol{k}|M_{j,\boldsymbol{k}}|\xi,\boldsymbol{k}\rangle\right],\label{eq:chi_G}\\
\bar{\chi}_{ij}^{(\text{R})}= & -\sum_{\xi\in\left\{ 1,2,3,4\right\} }\left[\int_{\text{FS}_{\xi}}\text{d}\theta\text{d}\phi\frac{k_{\text{F},\text{R}}^{2}(\theta,\phi)\text{sin}\theta}{v_{\text{F},\text{R}}(\theta,\phi)}\langle\xi,\boldsymbol{k}|M_{i,\boldsymbol{k}}|\xi,\boldsymbol{k}\rangle\langle\xi,\boldsymbol{k}|M_{j,\boldsymbol{k}}|\xi,\boldsymbol{k}\rangle\right],\label{eq:chi_R}
\end{align}
where $k_{\text{F},\Gamma(\text{R})}$ is the magnitude of the Fermi
momentum and $v_{\text{F},\Gamma(\text{R})}$ is the Fermi velocity
at the Fermi sheets around $\Gamma$ (R). For the numerical evaluation,
each numerical surface integration is done on $\theta\times\phi$
with uniform $100\times200$ grid points.

\subsection{$\Delta_{\pm}$, $I_{6(4)}$ and other superconducting gap functions }

Solving the LGE, we get a solution as a mixture of $\Delta_{\text{0-8}}$
which should be translated as $\Delta(\boldsymbol{k})=\sum_{i}\Delta_{i}M_{i,\boldsymbol{k}}$.
Note that $M_{\text{0-4},\boldsymbol{k}}$ are even under $\boldsymbol{k}\rightarrow-\boldsymbol{k}$
and constant at $\Gamma$ or R, while $M_{\text{5-8},\boldsymbol{k}}$
are odd under $\boldsymbol{k}\rightarrow-\boldsymbol{k}$ whose leading
order terms are linear in $\boldsymbol{k}$ around $\Gamma$ and R.
According to the parity under $\boldsymbol{k}\rightarrow-\boldsymbol{k}$,
we call $M_{\text{0-4},\boldsymbol{k}}$ and $M_{\text{5-8},\boldsymbol{k}}$
the even-parity pairings and the odd-parity pairings, respectively.
In the main text, we focus on the effects of $\Delta_{\pm}^{\Gamma(\text{R})}$,
thus only the even-parity parings that are reduced to either $I_{6(4)}$
or $\Delta_{\pm}^{\Gamma(\text{R})}$ are considered, while the contribution
from the odd-parity gap functions is neglected. To test the validity
of neglecting the odd-parity gap functions, we examine our numerical
results obtained from the LGE in detail by checking the surface Green's
function at every grid point in the phase diagram of Fig. 3(a) in
the main text.

\begin{figure}
\centering \includegraphics[width=0.95\textwidth]{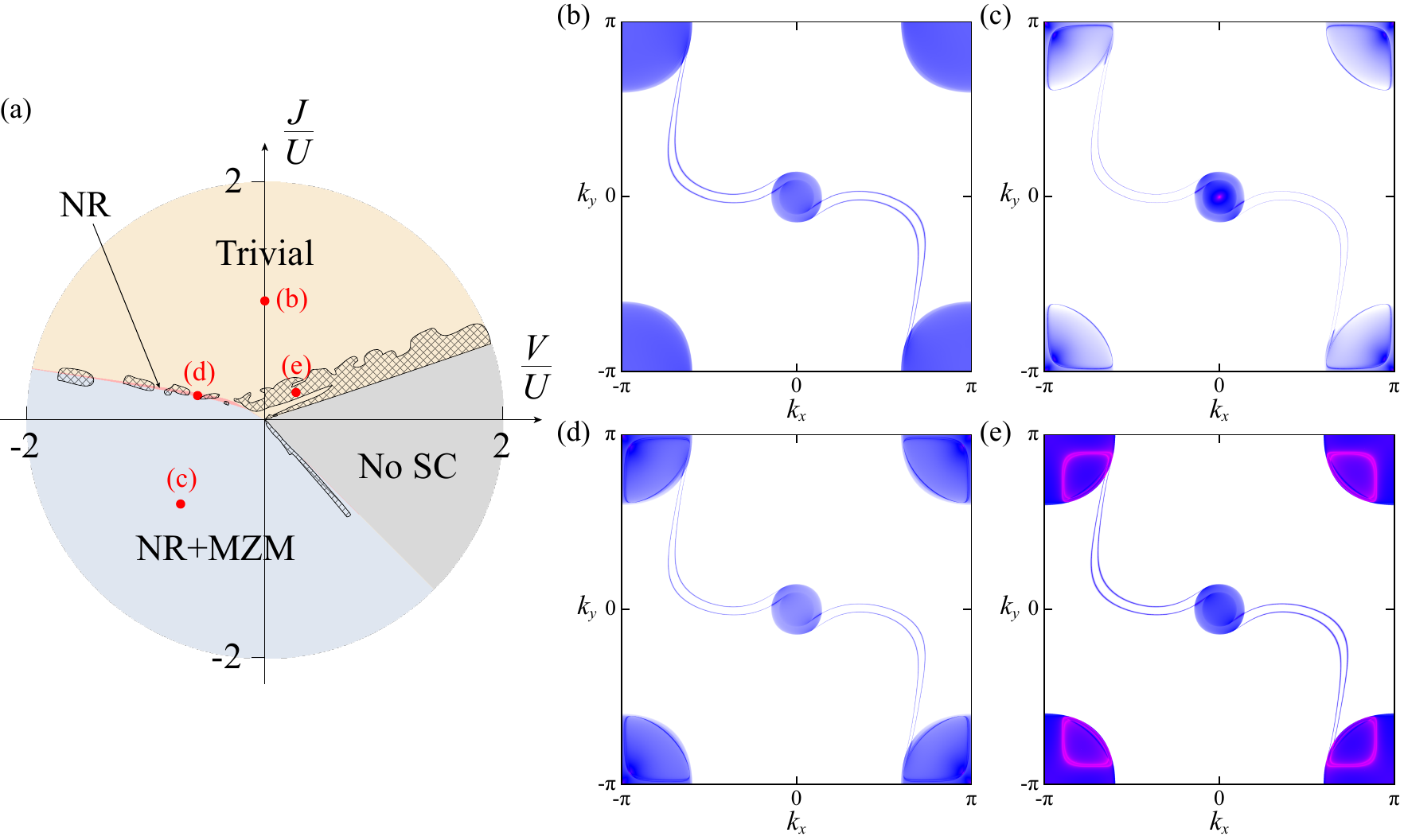} \caption{\label{fig:figS3}Phase diagram drawn by examining the surface Green's
function. The dashed regions represent the deviation from Fig. 3(a)
in the main text. In (b)-(e), the representative surface spectral
weights from the surface Green's function calculation are displayed.}
\end{figure}

Figure \ref{fig:figS3} summarizes our results. Figure \ref{fig:figS3}(a)
is the phase diagram in Fig. 3(a) in the main text over which four
representative points are marked by red dots. Here the dashed regions
represent the deviation from Fig. 3(a) where the solution of the LGE
is found to be not consistent with the description in the main text
using only $I_{6(4)}$ and $\Delta_{\pm}^{\Gamma(\text{R})}$. The
representative surface spectral weights from the surface Green's function
for the ``Trivial'' (beige), ``NR+MZM'' (blue), and ``NR'' (red)
regions are displayed in Figs. \ref{fig:figS3}(b), \ref{fig:figS3}(c),
and \ref{fig:figS3}(d), respectively. Figure \ref{fig:figS3}(e)
shows the surface spectral weight at the point (e) in Fig. \ref{fig:figS3}(a)
as a representative point in the dashed regions in Fig. \ref{fig:figS3}(a).
In those regions, the overall magnitude of $\sum_{i=\text{5-8}}\Delta_{i}M_{i,\boldsymbol{k}}$
at the Fermi sheets is not quite small compared to that of $\sum_{i=\text{0-4}}\Delta_{i}M_{i,\boldsymbol{k}}$.
The non-negligible odd-parity gap functions $\sum_{i=\text{5-8}}\Delta_{i}M_{i,\boldsymbol{k}}$
give rise to the unexpected nodal rings around R in Fig. \ref{fig:figS3}(e).
However, those regions are narrow and lie around the boundary between
the three major regions (``Trivial'', ``No SC'' and ``NR+MZM'').
Therefore, we conclude that the cases in which the odd-parity gap
functions play the major role are very limited, and thus it is sufficient
to consider the even-parity gap functions.

\subsection{Ambiguity of SOC parameters and their effect on the superconducting
gap functions }

So far, our discussion is based on the result from the tight-binding
model with $v_{s,i}=0$. However, as we discussed in Sec. \ref{sec:TB model for RhSi},
there is an ambiguity in choosing the $v_{s,i}$'s in the tight-binding
model. The presence of non-zero $v_{s,i}$ may distort the spin texture
on the Fermi sheets, and the pairing susceptibility $\chi$ is influenced
by these parameters. Though the topological nature does not change
much by the changes of $v_{s,i}$, the magnitude of the odd-parity
pairings can. These changes in the odd parity pairings whose leading
order terms are linear in $\bm{k}$ near the $\Gamma$ and R points
can induce deviations from the phase described by $\Delta_{\pm}$
and $I_{6(4)}$ which are constants in the leading order in $\bm{k}$.

\begin{figure}
\centering \includegraphics[width=0.95\textwidth]{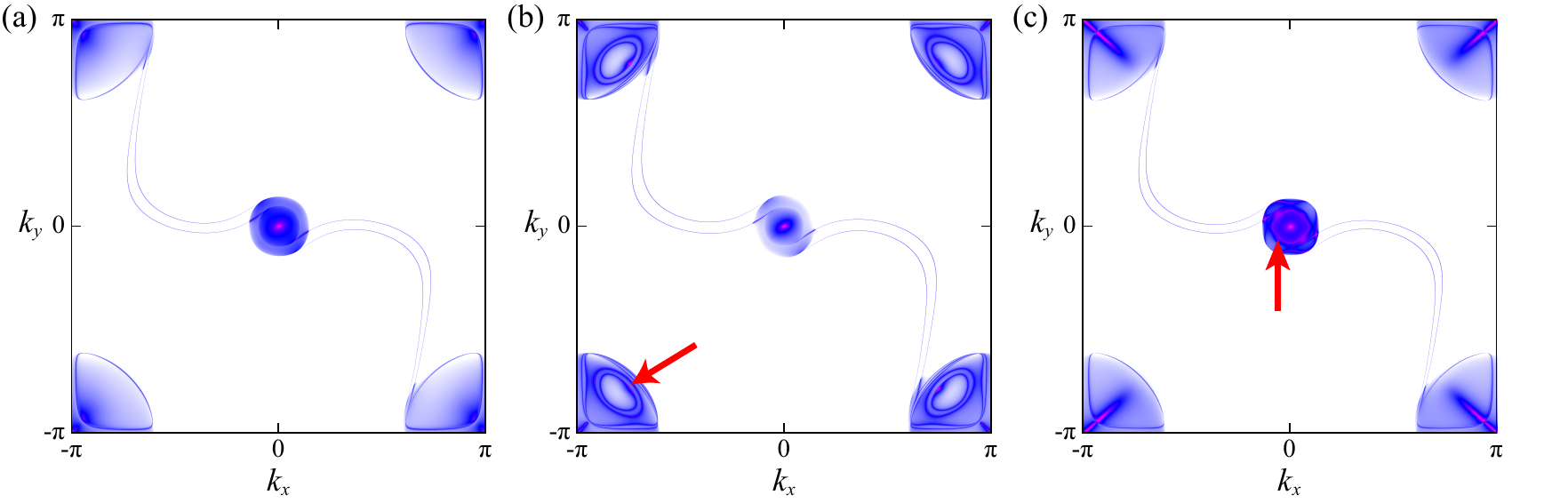} \caption{\label{fig:figS4} Surface Green's functions when $v_{s,1}$ is (a)
0, (b) $0.04$ eV, (c) $-0.04$ eV with $\mu=1.3m_{\Gamma}$, $U=1$,
and $V=J=-\frac{1}{5\sqrt{2}}$. Additional NRs are found at (b) R
and (c) $\Gamma$.}
\end{figure}

Figure \ref{fig:figS4} shows the surface Green's function calculations
for $v_{s,1}=0$, $0.04$, $-0.04$ eV with $v_{s,2}=v_{s,3}=0$ as
examples. As we change the SOC parameters, additional NRs can be found,
as indicated by the red arrows in Figs. \ref{fig:figS4}(b) and \ref{fig:figS4}(c).
Nonetheless, comparing Fig. \ref{fig:figS4}(a) with Figs. \ref{fig:figS4}(b)
and \ref{fig:figS4}(c), the topological features remain unaltered
such as the MZM at the $\bar{\Gamma}$ point and the NRs around the R point.

\bibliography{SMref}